%% file: sn-article.tex
\documentclass[pdflatex,sn-mathphys-num]{sn-jnl}% Math and Physical Sciences Numbered Reference Style

\usepackage{geometry} % 
\usepackage{graphicx}%
\usepackage{multirow}%
\usepackage{float}%
\usepackage{hyperref}%
\usepackage{amsmath,amssymb,amsfonts}%
\usepackage{amsthm}%
\usepackage{mathrsfs}%
\usepackage[title]{appendix}%
\usepackage{xcolor}%
\usepackage{textcomp}%
\usepackage{manyfoot}%
\usepackage{booktabs}%
\usepackage{algorithm}%
\usepackage{algorithmicx}%
\usepackage{algpseudocode}%
\usepackage{listings}%
\usepackage{tcolorbox}
\usepackage{listings}
\usepackage{xcolor}

\theoremstyle{thmstyleone}%
\theoremstyle{thmstyletwo}%

\theoremstyle{thmstylethree}%
\usepackage{tcolorbox}
\tcbuselibrary{breakable}

\begin{document}

\title[Article Title]{TO-Master: an LLM-agent framework for automated topology optimization}


\author[1]{\fnm{Haoju} 
\sur{Lin}}

\author[1]{\fnm{Wenchang}
\sur{Zhang}}

\author[1]{\fnm{Weipeng}
\sur{Xu}}

\author[2]{\fnm{Xiang}
\sur{Li}}

\author*[1]{\fnm{Tian} \sur{Xu}}\email{tianxu@ust.hk}

\author*[1]{\fnm{Tianju} \sur{Xue}}\email{cetxue@ust.hk}

\affil[1]{\orgdiv{Department of Civil and Environmental Engineering}, \orgname{The Hong Kong University of Science and Technology}, \orgaddress{\city{Hong Kong},  \country{China}}}

\affil[2]{\orgdiv{China Iron and Steel Research Institute Group},  \orgaddress{\city{Beijing}, \country{China}}}

% \affil[2]{\orgdiv{Department}, \orgname{Organization}, \orgaddress{\street{Street}, \city{City}, \postcode{10587}, \state{State}, \country{Country}}}

% \affil[3]{\orgdiv{Department}, \orgname{Organization}, \orgaddress{\street{Street}, \city{City}, \postcode{610101}, \state{State}, \country{Country}}}


\abstract{Topology optimization (TO) has become a mature computational design
method, but using it still requires substantial manual effort in geometry
preparation, mesh generation, boundary-condition assignment, solver setup, and
postprocessing. This implementation barrier limits the use of TO outside expert
workflows, even when differentiable finite element solvers are available. This
work introduces \textbf{TO-Master}, a large language model (LLM) agent framework that turns
finite-element-based TO into a conversational, tool-orchestrated workflow. From
natural language instructions and optional mesh, geometry, or image inputs, the
agent selects computational tools, constructs finite element TO models, checks
meshes and boundary conditions, and launches sensitivity-based optimization with
typed solver arguments. The framework supports generated and uploaded meshes,
image-to-mesh conversion, 2D and 3D structural compliance minimization, thermal
conduction, multiple load cases, stress-constrained optimization, and
engineering geometries. Numerical experiments show that TO-Master can reproduce
standard benchmark results and solve more complex engineering examples while
returning optimized results, field distributions, convergence histories, and
interactive artifacts without user-written code. An instruction ablation study
further shows that tool-usage rules, internal reasoning guidance, and few-shot
examples are critical for robust formulation under ambiguous user input. By
combining LLM-agent orchestration with deterministic finite element and
optimization tools, TO-Master removes the burden of trivial setup and routine
model construction, lowers the modeling barrier of TO, and preserves a reliable
numerical workflow. The TO-Master platform is available online at
\url{https://www.bohrium.com/en/apps/to-master}.}




\maketitle

\section{Introduction}

Topology optimization (TO) is a computational design method that optimizes the
material distribution in a prescribed design domain subject to boundary
conditions, constraints, and objective functions. Classical methods include the
homogenization-based method~\cite{BENDSOE1988197}, the solid isotropic material
with penalization (SIMP) method~\cite{ZHOU1991309,Rozvany1992,Bendsoe1999}, the
level-set method~\cite{WANG2003227,ALLAIRE2004363}, and the bi-directional
evolutionary structural optimization method~\cite{XIE1993885,Yang1999}. These
methods have been widely used for structural design, thermal design, and
multi-physics inverse design problems. However, setting up a TO problem remains
a nontrivial task. Users must specify geometries, meshes, boundary conditions, material models, objectives, constraints, optimization parameters, and
post-processing procedures. This process usually requires substantial
problem-specific scripting and repeated manual modification, which limits the
accessibility of TO workflows.

Recent advances in differentiable programming and automatic differentiation
provide new opportunities for computational mechanics and inverse design.
Automatic-differentiation-based TO frameworks such as \texttt{AuTO}~\cite{Chandrasekhar2021AuTO} or differentiable finite element method (FEM) frameworks such
as \texttt{JAX-FEM}~\cite{Xue2023JAXFEM} shows that finite element assembly,
solution procedures, and objective evaluations can be embedded in
automatic-differentiation-compatible computational graphs, making it possible to
obtain sensitivities through automatic differentiation. These developments reduce the implementation burden of
sensitivity analysis. Nevertheless, differentiable solvers still require users
to formulate the computational problem correctly. In practice, the main
difficulty often lies not only in solving the optimization problem constrained by partial differential equations (PDEs), but also in translating an engineering design intent into a complete
and valid solver input.

Large language models (LLMs) have shown strong capabilities in language
understanding, code generation, tool use, and multi-step reasoning.
Chain-of-thought prompting improves reasoning over intermediate steps
~\cite{wei2022cot}, while agentic frameworks such as ReAct enable LLMs to
interact with external tools during problem solving~\cite{yao2022react}.
Tool-augmented language models further demonstrate that LLMs can call external application programming interfaces (APIs) and use computational tools as part of their reasoning process
~\cite{schick2023toolformer}. These capabilities make LLM agents promising
interfaces for scientific computing workflows, where user intent must be mapped
to a sequence of structured computational operations. In computer-aided
engineering (CAE), LLM agents have been explored for LLM-empowered CAE
frameworks~\cite{guo2025llmcae}, automated structural
analysis~\cite{zhang2024structuralLLM}, computational fluid dynamics (CFD) case construction and solver
execution~\cite{fan2025chatcfd,pandey2025openfoamgpt}, and text-driven computer-aided design (CAD) generation~\cite{zhang2024llm4cad,wu2024texttodesign}. These studies show that
LLM agents can reduce manual intervention in simulation setup and workflow
orchestration. Recent and concurrent studies have also connected LLM agents with
topology optimization, including LLM-based controllers for adapting SIMP
continuation parameters~\cite{yang2026llmcontroller}, AutoSiMP for mapping
natural-language descriptions to SIMP topology optimization
problems~\cite{yang2026autosimp}, TO-Agents for preference-guided topology
optimization with multi-agent visual feedback~\cite{stewart2026toagents}, and
large-model-assisted conceptual structural design
realization~\cite{liang2025lmto}.

These works demonstrate the potential of LLM agents in TO, while two key challenges remain. First, many existing studies focus on relatively simplified structural
compliance benchmarks in academic settings, while more general problem settings,
engineering geometries, and user-level problem specifications remain less
explored. Second, because a valid TO workflow requires reliable
construction of boundary conditions, passive domains, physical constraints, and solver inputs, errors in these steps can produce plausible-looking but physically incorrect optimized structures.

\begin{figure}[H]
    \centering
    \includegraphics[width=1.0\linewidth]{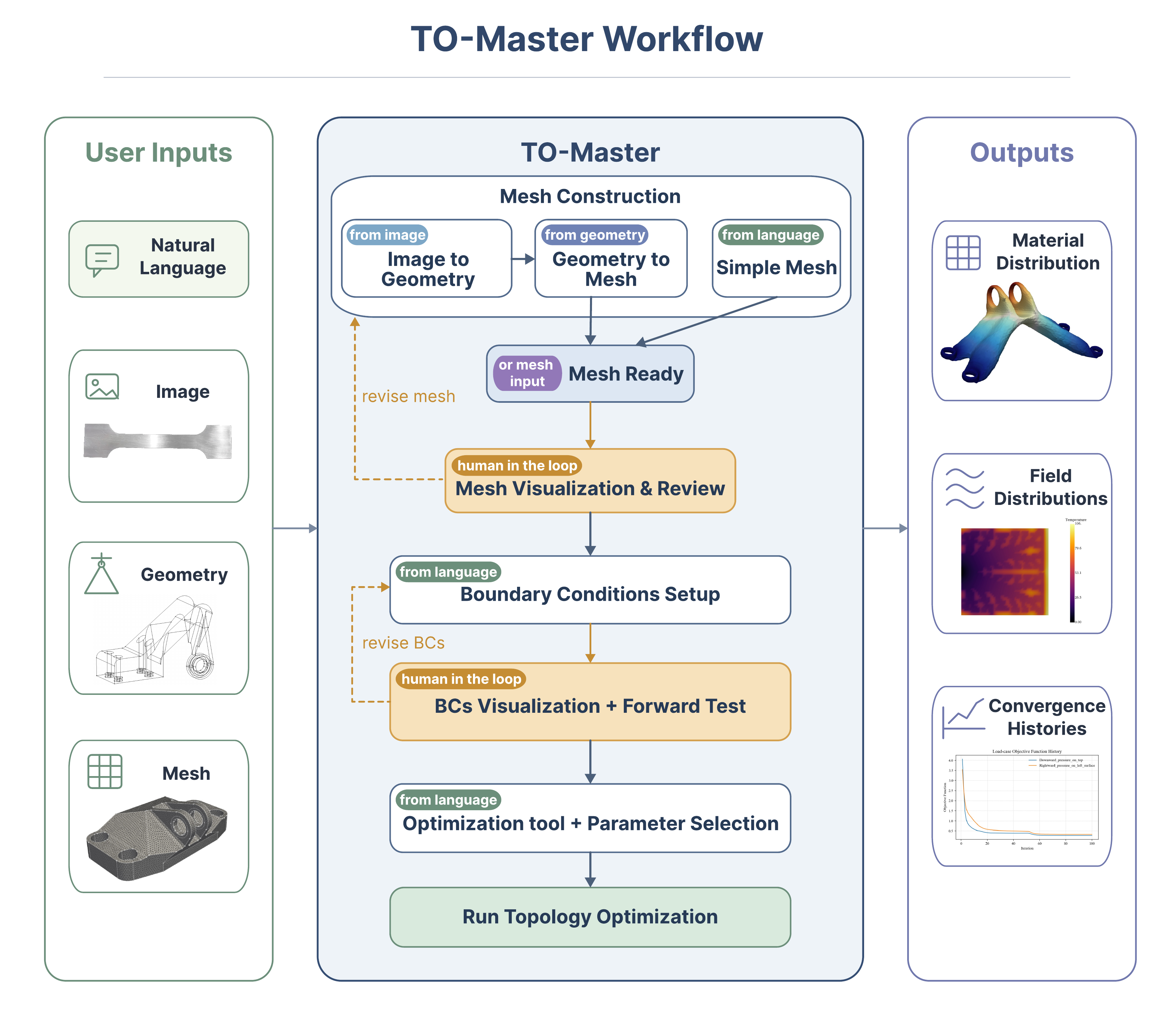}
    \caption{End-to-end workflow of the TO-Master framework. User-level natural
    language specifications are optionally supplemented by mesh files, geometry
    files, or image inputs, and are translated into FEM-based topology
    optimization problems. The framework builds or imports mesh models,
    visualizes and verifies meshes and boundary conditions through
    human-in-the-loop confirmation, performs sensitivity-based optimization, and
    returns optimized structures, field distributions, and convergence
    histories.}
    \label{fig:streamline}
\end{figure}

In this work, we present \textbf{TO-Master}, an LLM-agent framework for topology optimization. The main contributions of this work are reflected in three aspects: first, an LLM-agent-based workflow that translates user-level natural language specifications into finite-element TO models; second, a tool-orchestrated implementation that couples model construction, verification, optimization, and postprocessing through typed deterministic tools; and third, a systematic evaluation across benchmark, engineering, and prompt-ablation cases. TO-Master takes user-level natural language
specifications as input, optionally supplemented by mesh or geometry files, and supports 3D model and mesh generation from 2D image inputs. It
automatically formulates finite-element-based TO problems, visualizes and
verifies meshes and boundary conditions through human-in-the-loop confirmation,
performs sensitivity-based optimization with a differentiable FEM backend, and
reports optimized structures, field distributions, and convergence histories.
The framework supports 2D and 3D structural compliance minimization, thermal conduction, multiple load cases, and stress-constrained optimization for both regular design domains and arbitrary engineering geometries, including user-provided unstructured meshes and 3D geometries reconstructed from 2D images. By integrating structured prompting,
tool-level orchestration, human-in-the-loop verification, and
\texttt{JAX-FEM}-based sensitivity analysis, TO-Master provides a user-friendly
conversational interface for FEM-based topology optimization. The workflow is
shown in Fig.~\ref{fig:streamline}.

The remainder of this paper is organized as follows.
Section~\ref{sec:TO_formulation} presents the mathematical formulation of
density-based topology optimization. Section~\ref{sec:method} introduces the TO-Master framework, including the differentiable FEM backend and the LLM-agent design, which consists of the tool interface, prompt design, and agent workflow. Section~\ref{sec:examples} presents the benchmark and engineering examples. Section~\ref{sec:ablation} reports the prompt ablation study. Section~\ref{sec:discussion} discusses the effectiveness and robustness of the proposed framework. Finally,
Section~\ref{sec13} concludes the paper.

% In our previous work, we have developed jax-fem, but it still has some problem. auto-differentiation is very powerful.
% % Last paragraph
% In this work, we present \textbf{TO-Master}, a large-language-model--enabled
% framework that transforms topology optimization into a natural language--driven
% workflow. By integrating an LLM agent with a JAX-FEM-based differentiable
% optimizer, the proposed system interprets user-provided descriptions and optional
% geometric images, automatically constructs the corresponding finite element
% model, and executes the full optimization procedure without requiring any
% user-written code. This framework introduces a new mode of interaction for
% topology optimization, emphasizing accessibility, automation, and the reduction
% of repetitive modeling effort. Through a series of 2D and 3D examples involving
% regular and irregular geometries, we demonstrate that TO-Master provides a
% robust and flexible environment for conducting TO studies while preserving the
% accuracy and computational rigor of differentiable simulation-based optimization.

\section{Problem formulation for topology optimization}
\label{sec:TO_formulation}

This section summarizes the TO formulations implemented in TO-Master. We first
introduce the baseline density-based structural compliance minimization problem, and then
describe the density filtering, projection, passive-domain treatment, and
problem-specific extensions used in the implementation.

\subsection{Baseline structural compliance minimization}

The design domain \(\Omega\) is discretized into finite elements
\(\Omega_e\), \(e=1,\ldots,N_e\). In density-based TO, each element is assigned
a scalar density variable \(\rho_e\in[0,1]\), where \(\rho_e=0\) represents void
and \(\rho_e=1\) represents solid material. The design variables are collected
as \(\boldsymbol{\rho}=\{\rho_e\}_{e=1}^{N_e}\).

For structural optimization, the displacement field is approximated as
\begin{equation}
\boldsymbol{u}^h(\boldsymbol{x})
=
\sum_{k=1}^{N_n}
\phi_k(\boldsymbol{x})\boldsymbol{U}_k,
\label{eq:fe_displacement}
\end{equation}
where \(\phi_k\) is the finite element shape function associated with node
\(k\), \(\boldsymbol{U}_k\) is the nodal displacement vector, and \(N_n\) is the
number of nodes. The weak form is to find \(\boldsymbol{u}^h\in V_g\) such that
\begin{equation}
\int_{\Omega}
\boldsymbol{\sigma}(\boldsymbol{u}^h,\boldsymbol{\rho})
:
\boldsymbol{\varepsilon}(\boldsymbol{v})
\,d\Omega
=
\int_{\Gamma_t}
\boldsymbol{t}\cdot\boldsymbol{v}
\,d\Gamma,
\qquad
\forall \boldsymbol{v}\in V_0 ,
\label{eq:elastic_weak_form}
\end{equation}
where \(V_g\) satisfies the prescribed displacement boundary conditions,
\(V_0\) is the corresponding test space with homogeneous essential boundary
conditions, \(\boldsymbol{t}\) is the applied traction, and
\(\boldsymbol{\varepsilon}\) is the small-strain tensor.

The SIMP interpolation is used to relate density to stiffness
\cite{Bendsoe1999,BendsoeSigmund2003},
\begin{equation}
E_e =
E_{\min}
+
(E_{\max}-E_{\min})
\rho_e^p ,
\label{eq:simp_interpolation}
\end{equation}
where \(E_{\max}\) and \(E_{\min}\) are the solid and ersatz Young's moduli, and
\(p\) is the penalization factor.

The material usage is controlled by the standard volume-fraction constraint,
\begin{equation}
\frac{
\sum_{e=1}^{N_e}
\rho_e|\Omega_e|
}{
\sum_{e=1}^{N_e}
|\Omega_e|
}
\leq
V_f ,
\label{eq:volume_fraction_constraint}
\end{equation}
where \(|\Omega_e|\) denotes the element area in 2D or volume in 3D, and
\(V_f\) is the prescribed material volume fraction. In the implemented
structural optimization tools, this constraint is discretized as the mean
density over the optimized element set.

The baseline structural compliance minimization problem is written as
\begin{equation}
\begin{aligned}
\min_{\boldsymbol{\rho}} \quad
& J(\boldsymbol{\rho})
=
\boldsymbol{F}^{T}\boldsymbol{U},\\
\mathrm{s.t.}\quad
& \boldsymbol{K}(\boldsymbol{\rho})
\boldsymbol{U}
=
\boldsymbol{F},\\
& \frac{
\sum_{e=1}^{N_e}
\rho_e|\Omega_e|
}{
\sum_{e=1}^{N_e}
|\Omega_e|
}
\leq
V_f,\\
& 0\leq \rho_e\leq 1,
\qquad e=1,\ldots,N_e .
\end{aligned}
\label{eq:baseline_compliance_problem}
\end{equation}

\subsection{Density filtering, projection, and passive domains}

In the implementation, the raw density field is modified by density filtering,
optional Heaviside projection, and passive-domain treatment before finite
element analysis. Density filtering is used to regularize the design field and
reduce mesh-dependent numerical instabilities \cite{Bourdin2001}. When a problem
contains passive domains, only the elements outside the passive domains are
treated as designable. The set of designable elements is denoted by
\(\mathcal{I}_d\), and the passive elements are denoted by \(\mathcal{I}_p\).
The raw design variables are then
\(\boldsymbol{\rho}=\{\rho_e\}_{e\in\mathcal{I}_d}\).

The density filter is applied on the designable elements:
\begin{equation}
\bar{\rho}_e =
\frac{\sum_{j\in\mathcal{N}_e} w_{ej}\rho_j}
{\sum_{j\in\mathcal{N}_e} w_{ej}},
\qquad
w_{ej}=\max(0,r_{\min}-d_{ej}),
\label{eq:density_filter}
\end{equation}
where \(d_{ej}\) is the distance between element centroids and
\(\mathcal{N}_e\) is the set of neighboring designable elements within the
filter radius. The filter radius is defined as
\begin{equation}
r_{\min}=r_f\bar{h},
\label{eq:filter_radius}
\end{equation}
where \(r_f\) is the prescribed filter factor and \(\bar{h}\) is the average
element size. In the implementation, \(\bar{h}\) is computed from the average
cell measure over the full finite element mesh,
\begin{equation}
\bar{h}
=
\left(
\frac{1}{N_e}\sum_{e=1}^{N_e}|\Omega_e|
\right)^{1/d},
\label{eq:average_element_size}
\end{equation}
where \(d\) is the spatial dimension. Thus, \(r_f\) is a dimensionless factor controlling the filter radius relative to the mesh size.

When projection is enabled, the filtered density is further mapped to the
physical density field \(\tilde{\boldsymbol{\rho}}\) through a Heaviside-type
projection \cite{GuestPrevostBelytschko2004,WangLazarovSigmund2011Projection},
\begin{equation}
\tilde{\rho}_e =
\frac{\tanh(\beta\eta)+\tanh[\beta(\bar{\rho}_e-\eta)]}
{\tanh(\beta\eta)+\tanh[\beta(1-\eta)]},
\label{eq:heaviside_projection}
\end{equation}
where \(\beta\) controls the projection sharpness and \(\eta\) is the projection
threshold. If projection is not used, \(\tilde{\rho}_e=\bar{\rho}_e\).

The full element density field used in finite element analysis (FEA) is
\begin{equation}
\rho_e^{\mathrm{full}} =
\begin{cases}
\tilde{\rho}_e, & e\in\mathcal{I}_d,\\
1, & e\in\mathcal{I}_p .
\end{cases}
\label{eq:full_density}
\end{equation}
Therefore, \(\boldsymbol{\rho}\) is updated by the optimizer, while \(\tilde{\boldsymbol{\rho}}\) defines the physical densities on the designable elements. The assembled field \(\boldsymbol{\rho}^{\mathrm{full}}\) defines the material distribution over the whole mesh and is used in FEA and objective evaluation. For structural problems with passive domains, the structural tools apply the
implemented material constraint over the designable elements:
\begin{equation}
\frac{1}{|\mathcal{I}_d|}
\sum_{e\in\mathcal{I}_d}
\tilde{\rho}_e
\leq
V_f .
\label{eq:designable_volume_constraint}
\end{equation}

\subsection{Multi-load-case topology optimization}

For problems with multiple load cases, the same density field is shared by all
load cases. The objective is the sum of compliances,
\begin{equation}
J_{\mathrm{multi}}(\boldsymbol{\rho})
=
\sum_{m=1}^{N_L}
\boldsymbol{F}_m^{T}\boldsymbol{U}_m,
\qquad
\boldsymbol{K}(\boldsymbol{\rho}^{\mathrm{full}})
\boldsymbol{U}_m
=
\boldsymbol{F}_m ,
\label{eq:multi_load_objective}
\end{equation}
where \(N_L\) is the number of load cases. The material-volume constraint is imposed on the densities of the designable elements using Eq.~\eqref{eq:designable_volume_constraint}.

\subsection{Stress-constrained topology optimization}

For stress-constrained problems, TO-Master uses a differentiable aggregated
stress constraint. Directly constraining the maximum von Mises stress is
numerically inconvenient because the maximum operator is non-smooth and stresses
in low-density elements may introduce singular or noisy sensitivities. To obtain
a stable constraint for gradient-based optimization, the stress is evaluated
using a density-weighted stress measure and then aggregated by a \(p\)-norm,
following the maximum-stress-measure strategy used in stress-constrained TO
\cite{Yang2018StressConstraint}.

The density-weighted stress is computed with
\begin{equation}
E_e^{\sigma}
=
E_{\max}
\left(\rho_e^{\mathrm{full}}\right)^{p_{\sigma}},
\label{eq:stress_weighted_modulus}
\end{equation}
where \(p_{\sigma}=0.5\) in the implementation. This weighting reduces the
influence of low-density regions in the stress constraint while retaining a
smooth dependence on the design variables. The elemental stress value is defined
as the element-averaged density-weighted von Mises stress,
\begin{equation}
\bar{\sigma}_{\mathrm{vm},e}
=
\frac{
\int_{\Omega_e}
\sigma_{\mathrm{vm}}(\boldsymbol{x})\,d\Omega
}{
\int_{\Omega_e}d\Omega
}.
\label{eq:element_averaged_vm}
\end{equation}
The element average reduces quadrature-point noise and provides one
representative stress value for each element.

The global stress measure is approximated by
\begin{equation}
P_{\sigma}
=
\left[
\sum_{e}
\left(
\frac{
\bar{\sigma}_{\mathrm{vm},e}
}{
\sigma_{\mathrm{allow}}
}
\right)^{p_s}
\right]^{1/p_s},
\label{eq:stress_p_norm}
\end{equation}
where \(p_s=12\) and \(\sigma_{\mathrm{allow}}\) is the prescribed allowable
stress. To improve the approximation to the maximum stress, an
iteration-dependent scaling factor is applied \cite{Yang2018StressConstraint}:
\begin{equation}
c_{\sigma}^{(k)}
=
q_n
\frac{
\sigma_{\max}^{(k-1)}
}{
\sigma_{\mathrm{allow}}
P_{\sigma}^{(k-1)}
}
+
(1-q_n)c_{\sigma}^{(k-1)},
\qquad
q_n=0.5 .
\label{eq:stress_scaling_factor}
\end{equation}
The stress constraint is then written as
\begin{equation}
g_{\sigma}(\boldsymbol{\rho})
=
c_{\sigma}P_{\sigma}-1
\leq 0 .
\label{eq:stress_constraint}
\end{equation}
The complete stress-constrained problem is
\begin{equation}
\begin{aligned}
\min_{\boldsymbol{\rho}} \quad
& J(\boldsymbol{\rho})
=
\boldsymbol{F}^{T}\boldsymbol{U},\\
\mathrm{s.t.}\quad
& \boldsymbol{K}(\boldsymbol{\rho}^{\mathrm{full}})
\boldsymbol{U}
=
\boldsymbol{F},\\
& \frac{1}{|\mathcal{I}_d|}
\sum_{e\in\mathcal{I}_d}
\tilde{\rho}_e
\leq
V_f,\\
& g_{\sigma}(\boldsymbol{\rho})\leq 0,\\
& 0\leq \rho_e\leq 1,
\qquad e\in\mathcal{I}_d .
\end{aligned}
\label{eq:stress_problem}
\end{equation}
After optimization, TO-Master also evaluates the von Mises stress on the
thresholded solid topology for validation and visualization. This
post-processed solid-material stress is not used as the optimization constraint.

\subsection{Thermal topology optimization}

For thermal conduction problems, TO-Master solves the steady-state heat conduction equation in weak form: find \(T^h\in V_T\) such that
\begin{equation}
\int_{\Omega}
k(\rho^{\mathrm{full}})
\nabla T^h\cdot\nabla v
\,d\Omega
=
\int_{\Omega}
qv
\,d\Omega
+
\int_{\Gamma_q}
\bar{q}v
\,d\Gamma,
\qquad
\forall v\in V_0 ,
\label{eq:thermal_weak_form}
\end{equation}
where \(T^h\) is the finite element temperature field, \(q\) is the volumetric heat source, \(V_T\) satisfies the prescribed temperature boundary conditions on \(\Gamma_T\), \(V_0\) is the corresponding test space with homogeneous temperature conditions on \(\Gamma_T\), and \(\bar{q}\) is the prescribed heat flux on \(\Gamma_q\). Boundary segments without prescribed heat flux are treated as insulated boundaries, i.e., \(\bar{q}=0\). The effective thermal conductivity is interpolated as
\begin{equation}
k_e =
k_{\min}
+
(k_{\max}-k_{\min})
\left(\rho_e^{\mathrm{full}}\right)^p ,
\label{eq:thermal_conductivity}
\end{equation}
where \(k_{\max}\) and \(k_{\min}\) are the solid and ersatz conductivities.

The thermal objective is the heat-source-weighted temperature integral,
\begin{equation}
J_T(\boldsymbol{\rho})
=
\int_{\Omega}
qT\,d\Omega .
\label{eq:thermal_objective}
\end{equation}
This objective follows the standard thermal compliance-type formulation used in
heat-conduction TO \cite{GersborgHansen2006HeatConduction}. Unlike the
structural tools, the thermal optimization tools evaluate the volume-fraction
constraint using the element area or volume over the full domain, including
passive solid elements:
\begin{equation}
\frac{
\sum_{e\in\mathcal{I}_d}
\tilde{\rho}_e|\Omega_e|
+
\sum_{e\in\mathcal{I}_p}
|\Omega_e|
}{
\sum_{e}|\Omega_e|}
\leq
V_f .
\label{eq:thermal_volume_constraint}
\end{equation}
The thermal optimization problem is written as
\begin{equation}
\begin{aligned}
\min_{\boldsymbol{\rho}} \quad
& J_T(\boldsymbol{\rho}),\\
\mathrm{s.t.}\quad
& -\nabla\cdot
\left[
k(\rho^{\mathrm{full}})\nabla T
\right]
=
q,\\
& \frac{
\sum_{e\in\mathcal{I}_d}
\tilde{\rho}_e|\Omega_e|
+
\sum_{e\in\mathcal{I}_p}
|\Omega_e|
}{
\sum_{e}|\Omega_e|}
\leq
V_f,\\
& 0\leq \rho_e\leq 1,
\qquad e\in\mathcal{I}_d ,
\end{aligned}
\label{eq:thermal_problem}
\end{equation}
where passive thermal
regions remain fully conductive and contribute to the total material volume.

\subsection{Sensitivity analysis and gradient-based optimization}
\label{subsec:sensitivity_optimization}

The TO problems described above are solved as PDE-constrained optimization
problems. Following the discretize-then-optimize strategy used in
\texttt{JAX-FEM}~\cite{Xue2023JAXFEM}, the governing equations are first
discretized by FEM, and the resulting finite-dimensional optimization problem is
then solved. To cover both structural and thermal problems, the discrete state
variable is denoted by \(\boldsymbol{s}\), where
\(\boldsymbol{s}=\boldsymbol{U}\) for structural analysis and
\(\boldsymbol{s}=\boldsymbol{T}\) for thermal conduction. The discretized governing equation
is written as
\begin{equation}
    \boldsymbol{C}(\boldsymbol{s},\boldsymbol{\rho})=\boldsymbol{0},
    \label{eq:general_residual}
\end{equation}
where \(\boldsymbol{C}\) is the assembled residual after imposing the prescribed
boundary conditions.

The solution of Eq.~\eqref{eq:general_residual} defines the state as an implicit
function of the design, \(\boldsymbol{s}=\boldsymbol{s}(\boldsymbol{\rho})\).
The reduced objective and constraints are
\begin{equation}
    \widehat{J}(\boldsymbol{\rho})
    =
    J\!\left(\boldsymbol{s}(\boldsymbol{\rho}),\boldsymbol{\rho}\right),
    \qquad
    \widehat{g}_i(\boldsymbol{\rho})
    =
    g_i\!\left(\boldsymbol{s}(\boldsymbol{\rho}),\boldsymbol{\rho}\right),
    \label{eq:reduced_objective_constraints}
\end{equation}
and the reduced TO problem is
\begin{equation}
\begin{aligned}
    \min_{\boldsymbol{\rho}\in[0,1]^{N_\rho}}
    \quad & \widehat{J}(\boldsymbol{\rho}), \\
    \mathrm{s.t.}\quad
    & \widehat{g}_i(\boldsymbol{\rho}) \le 0,
    \qquad i=1,\ldots,N_g ,
\end{aligned}
\label{eq:reduced_TO_problem_section3}
\end{equation}
where \(N_\rho\) is the number of design variables and \(N_g\) is the number of inequality constraints.

For gradient-based optimization, the derivative of the reduced objective is
computed by differentiating the residual equation. The chain rule gives
\begin{equation}
    \frac{\mathrm{d}\widehat{J}}{\mathrm{d}\boldsymbol{\rho}}
    =
    \frac{\partial J}{\partial\boldsymbol{s}}
    \frac{\mathrm{d}\boldsymbol{s}}{\mathrm{d}\boldsymbol{\rho}}
    +
    \frac{\partial J}{\partial\boldsymbol{\rho}},
    \label{eq:Jhat_total_derivative}
\end{equation}
while differentiating
\(\boldsymbol{C}(\boldsymbol{s}(\boldsymbol{\rho}),\boldsymbol{\rho})
=\boldsymbol{0}\) yields
\begin{equation}
    \frac{\partial \boldsymbol{C}}{\partial \boldsymbol{s}}
    \frac{\mathrm{d}\boldsymbol{s}}{\mathrm{d}\boldsymbol{\rho}}
    +
    \frac{\partial \boldsymbol{C}}{\partial \boldsymbol{\rho}}
    =
    \boldsymbol{0}.
    \label{eq:residual_total_derivative}
\end{equation}
Since the number of design variables is typically large in TO, the adjoint form
is used for efficiency. Introducing the adjoint variable
\(\boldsymbol{\lambda}\), we solve
\begin{equation}
    \left(
    \frac{\partial \boldsymbol{C}}{\partial \boldsymbol{s}}
    \right)^{T}
    \boldsymbol{\lambda}
    =
    \left(
    \frac{\partial J}{\partial \boldsymbol{s}}
    \right)^{T},
    \label{eq:adjoint_equation}
\end{equation}
and obtain
\begin{equation}
    \frac{\mathrm{d}\widehat{J}}{\mathrm{d}\boldsymbol{\rho}}
    =
    \frac{\partial J}{\partial\boldsymbol{\rho}}
    -
    \boldsymbol{\lambda}^{T}
    \frac{\partial \boldsymbol{C}}{\partial\boldsymbol{\rho}} .
    \label{eq:adjoint_gradient}
\end{equation}
The same adjoint procedure is applied to state-dependent constraints, such as
the aggregated stress constraint.

In TO-Master, residual assembly, constitutive evaluation, state solution,
objective evaluation, and constraint evaluation are implemented through
\texttt{JAX-FEM}~\cite{Xue2023JAXFEM}. The adjoint method is implemented with customized differentiation rules, which
enable the automated computation of the required sensitivity terms. This avoids manual implementation
of case-specific sensitivity expressions and keeps the sensitivities consistent
with the discretized FEM model.

With the objective and constraint sensitivities available, the design variables are updated using the method of moving asymptotes (MMA)~\cite{Svanberg1987}. At each iteration, the FEM solver
evaluates the state field, objective, constraints, and gradients, and MMA updates
the density variables under the prescribed move limit and constraint settings.

\section{LLM-agent framework design}
\label{sec:method}

\begin{figure}[H]
    \centering
    \includegraphics[width=1.0\linewidth]{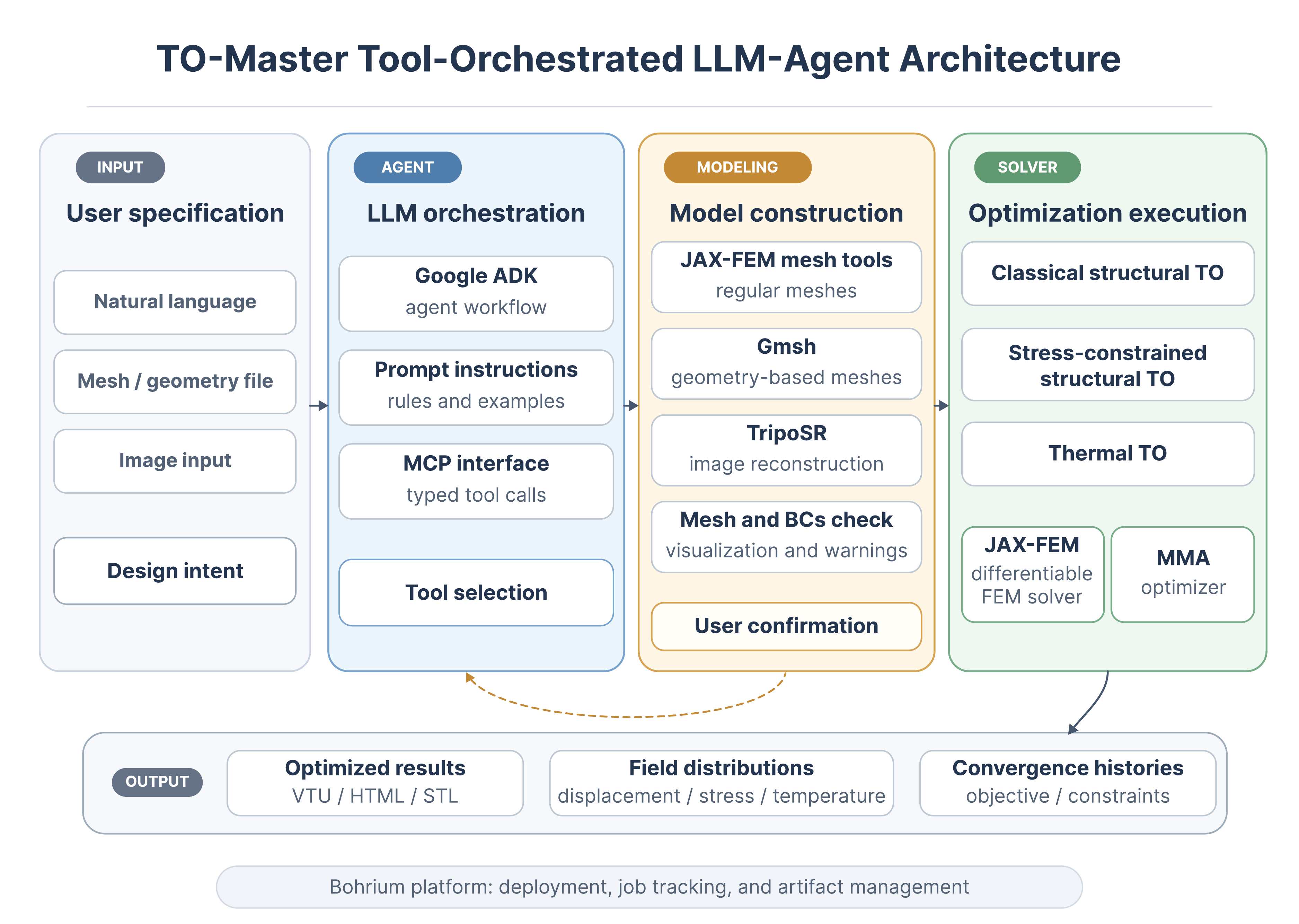}
    \caption{Overview of the TO-Master tool-orchestrated LLM-agent architecture. Natural language
    specifications and optional mesh, geometry, or image inputs are interpreted
    by the LLM agent and converted into verified finite element TO models. The
    framework supports model construction, mesh and boundary-condition checking,
    optimization execution, and the generation of optimized results, field
    distributions, and convergence histories.}
    \label{fig:flowchart}
\end{figure}

In this section, we present the LLM-agent framework design of
\textbf{TO-Master}. The numerical optimization follows the
sensitivity analysis and MMA-based update scheme described in
Sec.~\ref{subsec:sensitivity_optimization}. Built on the agent execution
infrastructure provided by Bohrium~\cite{zhang2025bohrium+}, TO-Master connects
user-level problem descriptions with finite-element-based TO tools. As shown in
Fig.~\ref{fig:flowchart}, the framework accepts natural language specifications
together with optional mesh, geometry, or image inputs. These inputs are
translated into a finite element TO model, whose mesh and boundary-condition
setup are checked before optimization. The framework then executes the
corresponding TO workflow and returns optimized results, field distributions,
and convergence histories.

The software stack is organized as follows. Google Agent Development Kit (ADK)~\cite{google2026adk} runs the LLM agent and manages dialogue state,
instructions, and tool selection. The Model Context Protocol (MCP)~\cite{modelcontextprotocol2026} defines
the typed interface through which the agent calls computational tools. The
\texttt{bohr-agent-sdk} handles tool execution, status tracking, logging, and
artifact management. The TO-Master MCP server implements the domain-specific TO
tools, including mesh processing, boundary-condition checking, topology
optimization, and result postprocessing. The FEA and sensitivity
evaluation are performed by the \texttt{JAX-FEM}-based solver described in
Sec.~\ref{subsec:sensitivity_optimization}.

As an LLM-agent framework, TO-Master converts user intent into structured tool
arguments and coordinates the execution of the TO workflow. Its numerical
reliability comes from coupling this agent-level coordination with deterministic
computational tools: mesh generation, mesh inspection, boundary-condition
visualization, FEA, sensitivity evaluation, optimization, and postprocessing are
performed by predefined solver modules. Thus, TO-Master provides a
conversational interface for TO problem formulation and execution while
preserving the underlying numerical modeling and optimization procedures.

The following subsections describe the main framework components of TO-Master: the few-shot and reasoning-guided prompt design, the TO-Master tool layer, and the LLM-agent-orchestrated workflow.

\subsection{Few-shot and reasoning-guided prompt design}

The TO-Master agent is configured by a structured system instruction, whose
organization is shown in Fig.~\ref{fig:prompt}. The complete instruction used in
the numerical experiments is provided in Appendix~\ref{secA1}. The instruction contains four
parts: role specification, tool-usage rules, internal reasoning guidance, and
few-shot examples. 

The role specification defines the agent as a
domain-specialized assistant for JAX-FEM mesh generation, mesh and boundary-condition checking, and topology optimization. This role definition anchors the agent to TO-related tasks and prevents it from treating the interaction as a general code-generation problem.

The tool-usage rules define the allowed computational interface. They specify
the available MCP tools, legal boundary-condition selection formats, expected
argument fields, returned artifact structures, and user-confirmation
requirements. In particular, topology optimization tools require prior mesh and
boundary-condition checking, followed by user confirmation through the
\texttt{bc\_confirmed} flag. These rules constrain the execution order and
prevent the agent from directly launching an optimization run before the finite
element TO model has been checked.

The internal reasoning guidance provides a formulation checklist for the agent.
It guides the agent to identify the problem type, distinguish mechanical and
thermal TO, determine the spatial dimension, identify the mesh source, extract
boundary conditions, loads, passive domains, material parameters, volume
fraction, and optimization settings, and then select the next safe tool call.
This reasoning process is used internally and is not shown to the user; the user
only receives the resulting permission request, tool output summary, and links
to generated artifacts.

The few-shot examples provide concrete mappings from ambiguous user requests to
complete tool plans. They cover uploaded mesh preview, image-to-mesh conversion,
mechanical compliance TO, stress-constrained TO, thermal TO, multi-load-case TO,
and unsupported direct-solve requests. Each example includes an easy prompt, an
inferred complete specification, a concrete tool plan, and a user-visible
permission request. These examples help the agent resolve common ambiguities,
such as finite load patches, passive-domain assignment, insulated thermal
boundaries, and the choice between single-load, multi-load, and
stress-constrained solvers.

\begin{figure}[H]
    \centering
    \includegraphics[width=1.0\linewidth]{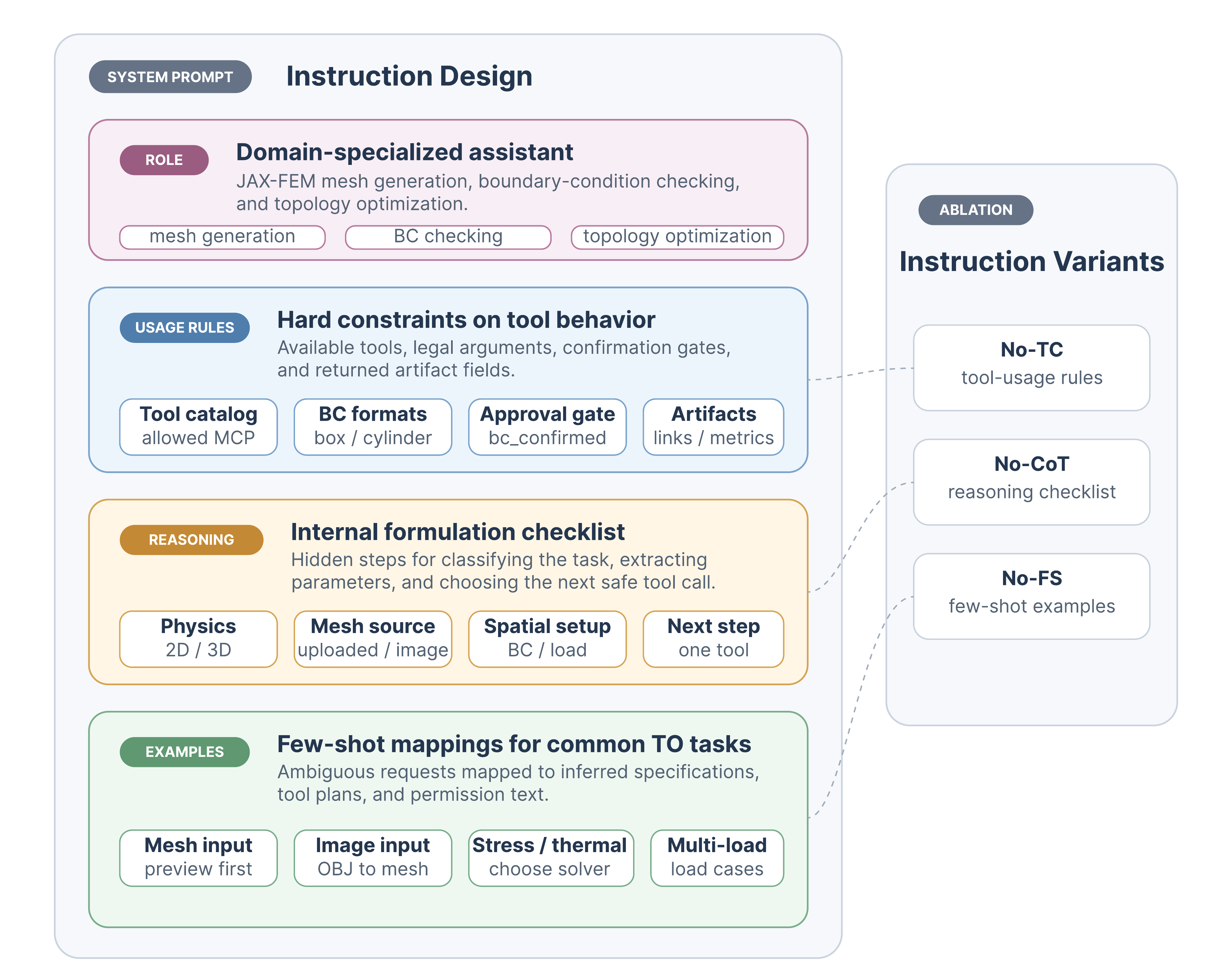}
    \caption{Structure of the TO-Master system instruction. The four modules
    define the agent's domain role, tool constraints, internal formulation
    checklist, and few-shot mappings, and correspond to the instruction
    ablations evaluated in this work.}
    \label{fig:prompt}
\end{figure}

\subsection{TO-Master tool layer}

The TO-Master MCP server exposes domain-specific tools for mesh preparation, mesh and boundary-condition checking, TO execution, and
postprocessing. Each tool is called with typed arguments and returns structured
outputs, including execution status, input parameters, artifacts, metrics, and
text summaries. This tool layer is central to TO-Master: it allows the LLM agent
to automate TO workflows by orchestrating deterministic numerical modules,
rather than by generating solver code.

For mesh preparation, TO-Master supports generated meshes, uploaded
meshes, and image-derived mesh models. Rectangular and box meshes are generated
using \texttt{JAX-FEM}-based utilities, while cylinder and geometry-based box
meshes can be generated using Gmsh~\cite{geuzaine2009gmsh}. Uploaded meshes are
inspected before use, with dimensions, bounding boxes, node and cell counts, and
preview images reported to the user. For image inputs, a TripoSR-based
module~\cite{TripoSR2024} reconstructs an OBJ surface, which is then converted
into a voxelized hexahedral volume mesh. This enables the same agent workflow to
handle regular benchmark domains, user-provided engineering meshes, and
image-derived geometry inputs.

For mesh and boundary-condition checking, TO-Master visualizes supports, loads,
and thermal boundary conditions selected by box or cylindrical regions. The
checking tools report empty selections and return preview images and interactive
HTML files. For mechanical problems, a full-material FEA check is also used to
detect nearly trivial displacement responses, which may indicate ineffective
loads or ill-posed boundary conditions. Multiple-load cases are checked case by case. 

With prior checking and user confirmation through the
\texttt{bc\_confirmed} flag, which reduce the risk of silent formulation errors, the optimization tools support 2D and 3D structural compliance TO, multi-load-case structural compliance TO, stress-constrained structural TO, and steady-state thermal conduction TO. They share common arguments for mesh paths,
boundary conditions, material parameters, volume fractions, filter settings,
move limits, iteration numbers, optional Heaviside projection, and passive
domains. Passive domains are passed as \texttt{undesign\_domains} and kept
solid during optimization. Postprocessing exports optimized density fields,
field distributions, convergence histories, VTU files, interactive HTML views,
and STL surface models for 3D optimized results, making both intermediate and
final results inspectable.

\subsection{LLM-agent-orchestrated workflow}

The agent workflow converts a user request into a checked and executable TO
problem. The LLM agent interprets the design domain, mesh source, boundary
conditions, loads, passive domains, material parameters, objective type,
physical constraints, and optimization settings, and then selects the
appropriate mesh, checking, and optimization tools. Regular domains are
generated directly, uploaded meshes are previewed before use, and image inputs
are processed through the image-to-geometry and geometry-to-mesh pipeline.

A key feature is staged user confirmation. Before critical tool calls, the agent
summarizes the inferred action, names the tool to be called, lists the key
parameters, and asks for user approval. After mesh and boundary-condition
checking, the agent reports artifacts, metrics, and warnings, and proceeds to
optimization only after confirmation. If a tool returns a warning or failure
status, the agent reports the issue and pauses instead of continuing silently.

After confirmation, the selected optimization tool is called with typed
arguments. The deterministic solver evaluates the finite element state, objective,
constraints, and sensitivities, and MMA updates the density variables as
described in Sec.~\ref{subsec:sensitivity_optimization}. After convergence,
TO-Master returns optimized results, field distributions, convergence histories,
and interactive visualizations. In this workflow, the LLM handles
interpretation, tool selection, and user interaction, while deterministic tools
handle mesh processing, FEA, optimization, and postprocessing. This
division makes the workflow reproducible while retaining the flexibility of
natural language interaction.

\section{Benchmark and engineering examples}
\label{sec:examples}

This section evaluates the robustness and applicability of TO-Master through a
set of benchmark and engineering topology optimization examples. The examples
cover two- and three-dimensional benchmark structures, thermal conduction,
multi-load structural compliance minimization, stress-constrained problems,
arbitrary geometries, and image-to-optimization cases. They are designed to
assess whether TO-Master can translate natural language specifications into
solver-ready topology optimization problems across different problem settings.

All examples use the same differentiable FEM backend and gradient-based
optimization framework. The geometry, boundary conditions, and optimization
constraints are varied across cases. For each example, the constructed boundary
conditions, convergence histories, constraint values, and final optimized
topologies are recorded for analysis.

Unless otherwise stated, all numerical values are interpreted in a consistent
unit system. In the International System of Units (SI) interpretation used for the regular structural benchmark cases, lengths are measured in \(\mathrm{m}\), forces in \(\mathrm{N}\), surface
tractions, Young's modulus, and von Mises stresses in \(\mathrm{Pa}\), and displacements in \(\mathrm{m}\). For thermal problems, temperature is reported in degrees Celsius
\((^\circ\mathrm{C})\), thermal conductivity in
\(\mathrm{W\,m^{-1}K^{-1}}\), and the internal heat source in
\(\mathrm{W\,m^{-3}}\), with the two-dimensional heat-conduction case treated as
a unit-thickness model. Since the steady-state heat-conduction problem depends
on temperature differences, using \(^{\circ}\mathrm{C}\) instead of
\(\mathrm{K}\) only changes the reference offset and does not affect the
optimization result. The solver does not
perform unit conversion; therefore, all input quantities are used directly in a
self-consistent unit system. In the natural language prompts, terms such as
``pressure'' are interpreted by the solver as surface tractions.

The eleven cases are organized into regular benchmark cases and extended
engineering examples. Cases 1--6 test standard structural, thermal, and
multi-load formulations on regular domains, whereas Cases 7--11 examine stress
constraints, arbitrary engineering geometries, and image-to-optimization.

Table~\ref{tab:benchmark_cases} lists the eleven benchmark cases with their geometries and boundary conditions. Cases 1--6 are regular geometry problems, including two-dimensional and
three-dimensional structural compliance minimization, thermal conduction, and a multi-load
tower structure. Cases 7--8 are stress-constrained optimization problems.
Cases 9--10 involve arbitrary three-dimensional engineering geometries. Case 11 uses a geometry and mesh reconstructed from a 2D image input.

All cases are formulated through natural language interaction with TO-Master,
without user-written solver code. The detailed prompts are provided in
Appendix~\ref{secA2}. The corresponding boundary conditions are shown in
Figs.~\ref{fig:TO_BC} and~\ref{fig:TO_BC2}, and the optimized results are shown
in Figs.~\ref{fig:TO_Result_full} and~\ref{fig:TO_Result_2}.

\subsection{Optimization on regular benchmark problems}

\begin{figure}[H]
    \centering
    \includegraphics[width=1\linewidth]{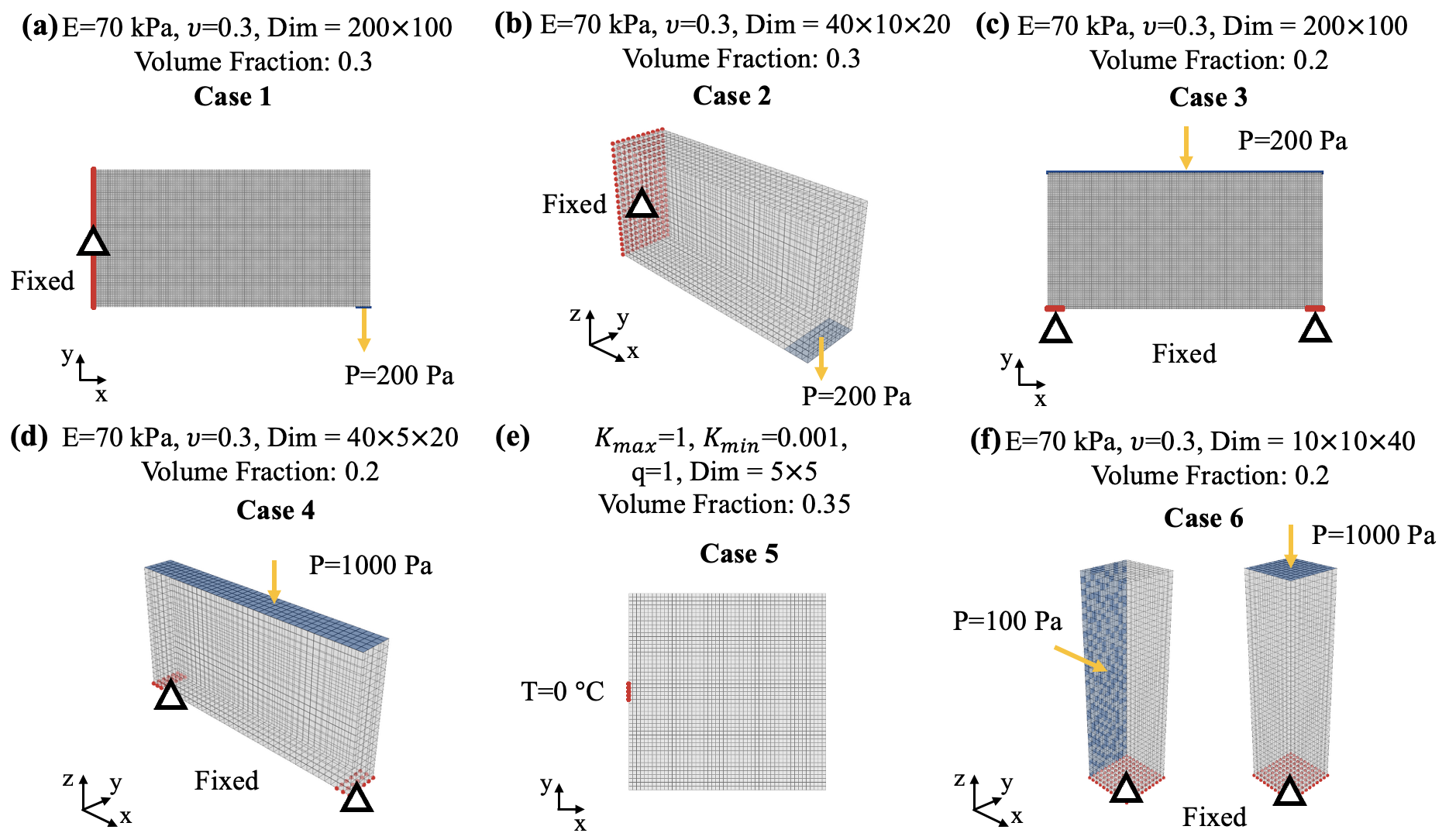}
    \caption{Boundary condition settings of topology optimization for different cases. (a) Case~1: 2D cantilever beam; (b) Case~2: 3D cantilever beam; (c) Case~3: 2D bridge; (d) Case~4: 3D bridge; (e) Case~5: 2D heat conduction; (f) Case~6: 3D tower with multiple load cases.}
    \label{fig:TO_BC}
\end{figure}

\begin{figure}[H]
    \centering
    \includegraphics[width=0.95\linewidth]{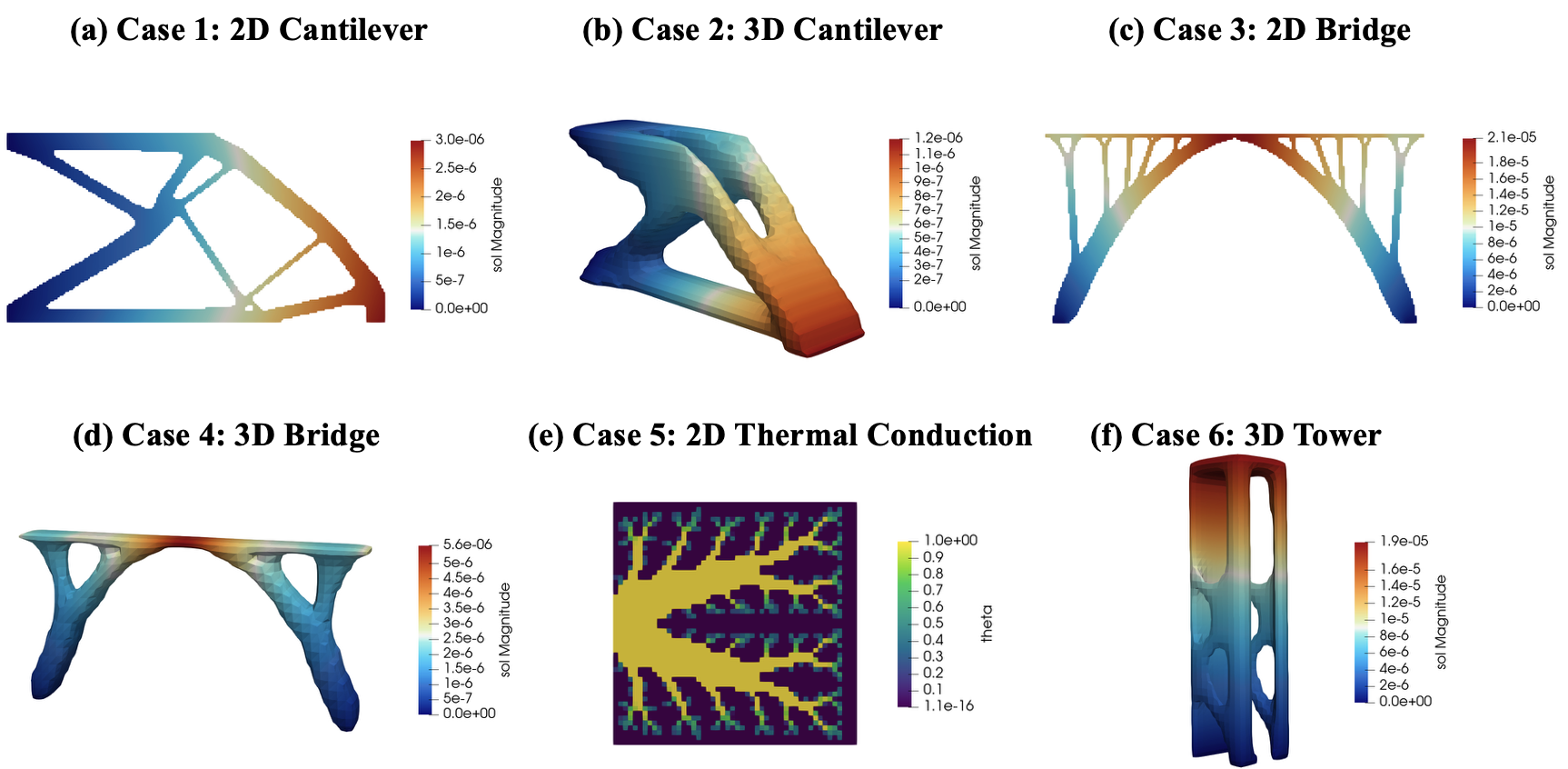}
    \caption{Topology optimization results for the benchmark cases. For each case, the figure depicts the displacement magnitude or the virtual density of the optimized topology under the prescribed boundary conditions. (a) Case~1: 2D cantilever beam; (b) Case~2: 3D cantilever beam; (c) Case~3: 2D bridge; (d) Case~4: 3D bridge; (e) Case~5: 2D heat conduction; (f) Case~6: 3D tower with multiple load cases.}
    \label{fig:TO_Result_full}
\end{figure}

Cases 1--6 in Table~\ref{tab:benchmark_cases} evaluate the basic optimization
capability of TO-Master on regular domains. Their boundary conditions are shown
in Fig.~\ref{fig:TO_BC}, and the corresponding optimized results are shown in
Fig.~\ref{fig:TO_Result_full}. These cases include cantilever beams, bridge
structures, a thermal conduction problem, and a multi-load tower problem.

The thermal benchmark, Case~5, uses a square domain with dimensions \(5\,\mathrm{m} \times 5\,\mathrm{m}\) and an element size of
\(0.1\,\mathrm{m}\). A uniform internal heat source of
\(1.0\,\mathrm{W\,m^{-3}}\) is applied over the whole domain. The cold sink is prescribed as a reference temperature
\(T=0\,^\circ\mathrm{C}\) on the middle part of the left boundary. The maximum and minimum thermal conductivities are
\(k_{\max}=1.0\,\mathrm{W\,m^{-1}K^{-1}}\) and
\(k_{\min}=10^{-3}\,\mathrm{W\,m^{-1}K^{-1}}\), respectively. The other
boundaries are insulated by omitting heat-flux boundary conditions.

For clarity, the intermediate interaction process is illustrated using the
three-dimensional bridge case as a representative example
(Case~4 in Table~\ref{tab:benchmark_cases}; see
Figs.~\ref{fig:TO_BC}(d) and~\ref{fig:TO_Result_full}(d)). The design domain has
dimensions \(40\,\mathrm{m} \times 5\,\mathrm{m} \times 20\,\mathrm{m}\). The
lower-left and lower-right corner regions are fixed with zero displacement, and
a downward surface traction of \(1000\,\mathrm{Pa}\) is applied on the top
surface. The top layer is preserved as a passive domain to maintain the
loading interface during optimization. The default Young's modulus is
\(E=70\,\mathrm{GPa}\), the Poisson's ratio is \(\nu=0.3\), and a volume fraction
constraint of 0.2 is imposed. The optimization is run for 100 iterations.

A fuzzy natural language prompt used for this case is shown below. This prompt
is the same as the fuzzy prompt used in the ablation study. The corresponding
coordinate-level precise prompt is provided in Appendix~\ref{secA2}.

\begin{tcolorbox}[title={Fuzzy natural language problem description used for Case 4},
                  colback=gray!5,
                  colframe=black!60,
                  fonttitle=\bfseries,
                  sharp corners]
\small
Create a 3D bridge benchmark with a 40 x 5 x 20 mesh. Support both lower end
corners and apply downward pressure of 1000 across the top surface. Keep the top
layer as a passive domain. Use 20 percent material, run 100 optimization steps,
set the step size to 0.1, and use a filter factor of 2.
\end{tcolorbox}

\begin{table}[!htbp]
\centering
\caption{Summary of benchmark topology optimization cases used for evaluating
the basic performance of TO-Master.}
\label{tab:benchmark_cases}
\begin{tabular}{c c c}
\hline
Case ID & Geometry & Boundary Conditions  \\
\hline
1 & 2D Cantilever Beam & Left edge fixed; \\&&surface traction applied at the lower-right corner. \\\\

2 & 3D Cantilever Beam & Left side fixed; \\&&surface traction applied at the lower-right corner. \\\\

3 & 2D Bridge & Bottom-left and bottom-right corners fixed;\\
  && surface traction applied on the top boundary. \\\\

4 & 3D Bridge & Bottom-left and bottom-right corners fixed;\\
  && surface traction applied on the top boundary. \\\\

5 & 2D Thermal Conduction & Uniform internal heat source; fixed-temperature cold\\
  && sink applied to the central region of the left boundary. \\\\

6 & 3D Tower (Multiple Load Cases) & Bottom fixed; two load cases considered: \\
  && 1) downward surface traction applied on the top surface; \\
  && 2) rightward surface traction applied on the left surface. \\\\

7 & 3D L-Shaped Beam (Stress Constraint) & Top-left support region fixed;\\
  && downward surface traction applied on \\&&the right-side loading patch;\\
  && loading patch preserved as a passive domain. \\\\

8 & 3D T-Shaped Beam (Stress Constraint) & Two support regions fixed;\\
  && downward surface traction applied on \\&&the central loading patch;\\
  && loading patch preserved as a passive domain. \\\\

9 & 3D Jet Engine Bracket (Arbitrary Geometry) & Four lower bolt holes fixed;\\
  && inclined surface traction applied at the upper pin hole;\\
  && bolt holes and pin hole preserved as passive domains. \\\\

10 & 3D Airplane Bearing Bracket (Arbitrary Geometry) & Four lower bolt holes fixed;\\
  && downward surface traction applied at \\&&the right-side pin hole;\\
  && bolt holes and pin hole preserved as passive domains. \\\\

11 & 3D Dog-Bone-Shaped Specimen (Image Input) & Left end fixed;\\
  && downward surface traction applied at the right end;\\
  && both end regions preserved as passive domains. \\

\hline
\end{tabular}
\end{table}

Using the full instruction, TO-Master translates the fuzzy natural language
description into a solver-ready problem, including mesh generation, boundary
condition visualization, passive-domain assignment, and topology optimization.

The optimized topologies exhibit clear load paths and structural features
consistent with classical topology optimization benchmarks, confirming that the constructed finite element
models and optimization settings are physically meaningful.

\subsection{Stress-constrained optimization}

\begin{figure}[H]
    \centering
    \includegraphics[width=1\linewidth]{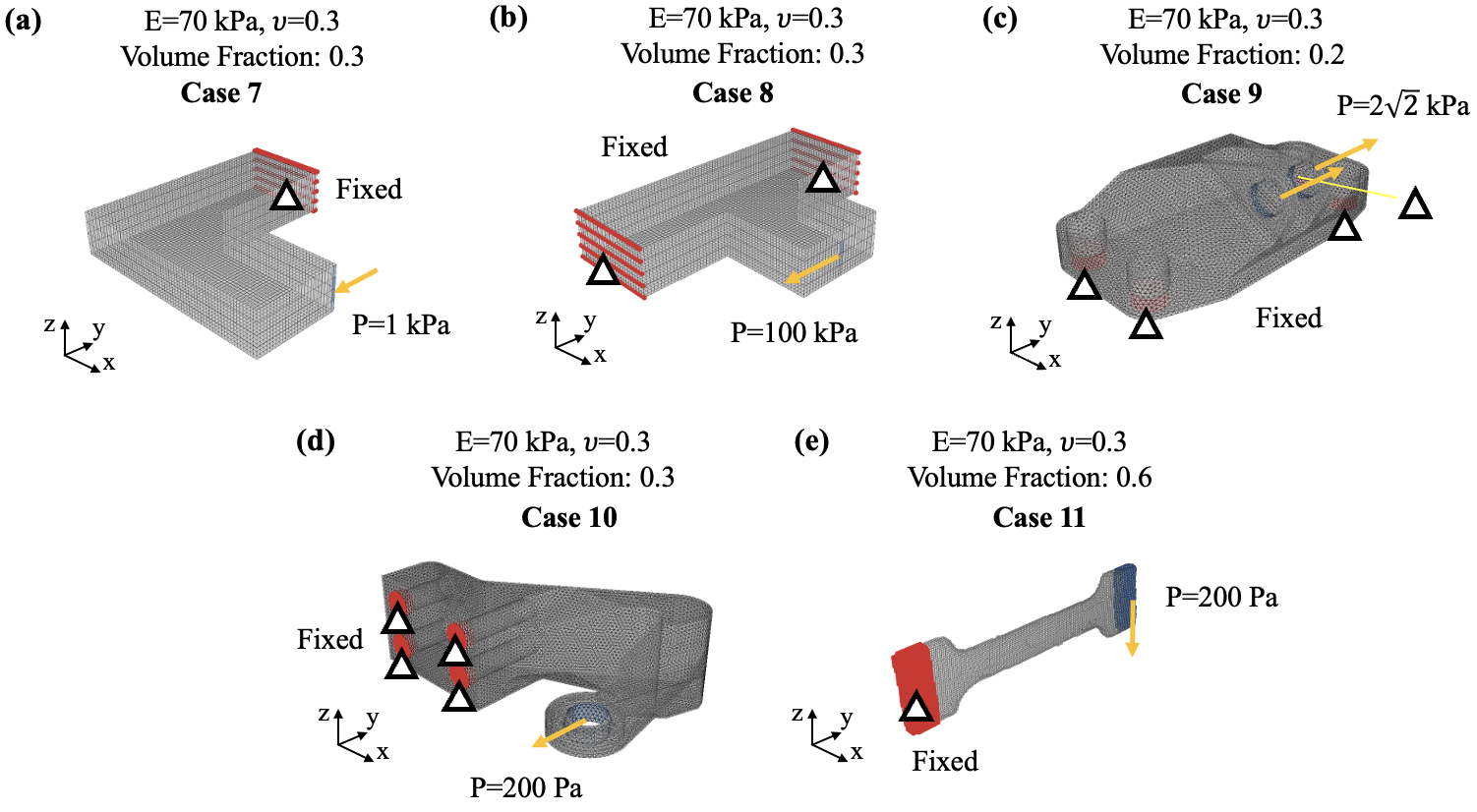}
    \caption{Boundary condition settings of topology optimization for different cases. (a) Case~7: 3D L-shaped beam with stress constraint; (b) Case~8: 3D T-shaped beam with stress constraint; (c) Case~9: 3D jet engine bracket with arbitrary geometry; (d) Case~10: 3D airplane bearing bracket with arbitrary geometry; (e) Case~11: 3D dog-bone-shaped specimen generated from image input.}
    \label{fig:TO_BC2}
\end{figure}

\begin{figure}[H]
    \centering
    \includegraphics[width=0.95\linewidth]{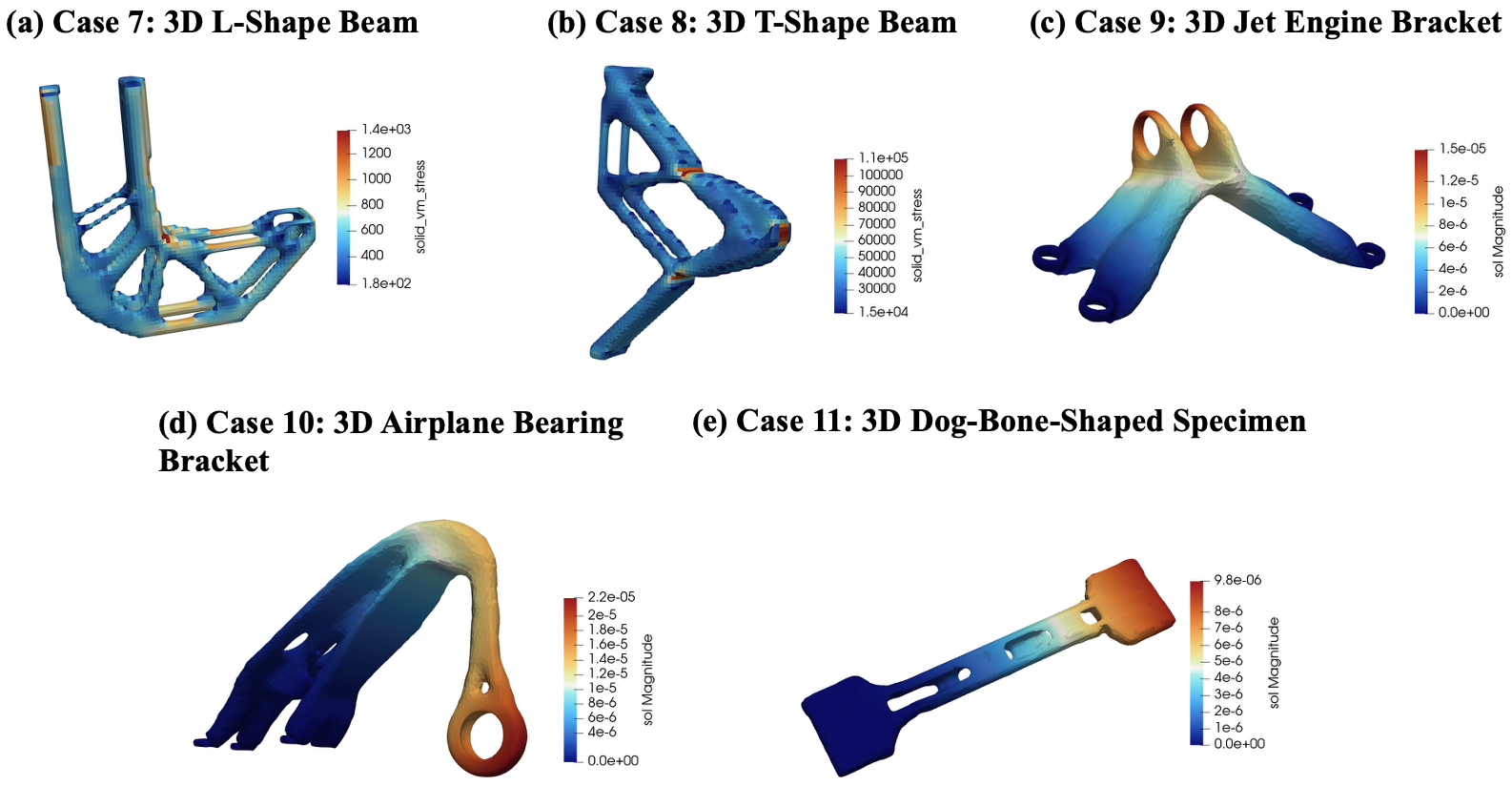}
    \caption{Topology optimization results for the benchmark cases. For each case, the figure depicts the von Mises stress or the displacement magnitude of the optimized topology under the prescribed boundary conditions. (a) Case~7: 3D L-shaped beam with stress constraint; (b) Case~8: 3D T-shaped beam with stress constraint; (c) Case~9: 3D jet engine bracket with arbitrary geometry; (d) Case~10: 3D airplane bearing bracket with arbitrary geometry; (e) Case~11: 3D dog-bone-shaped specimen generated from image input.}
    \label{fig:TO_Result_2}
\end{figure}

\begin{figure}[H]
    \centering
    \includegraphics[width=0.95\linewidth]{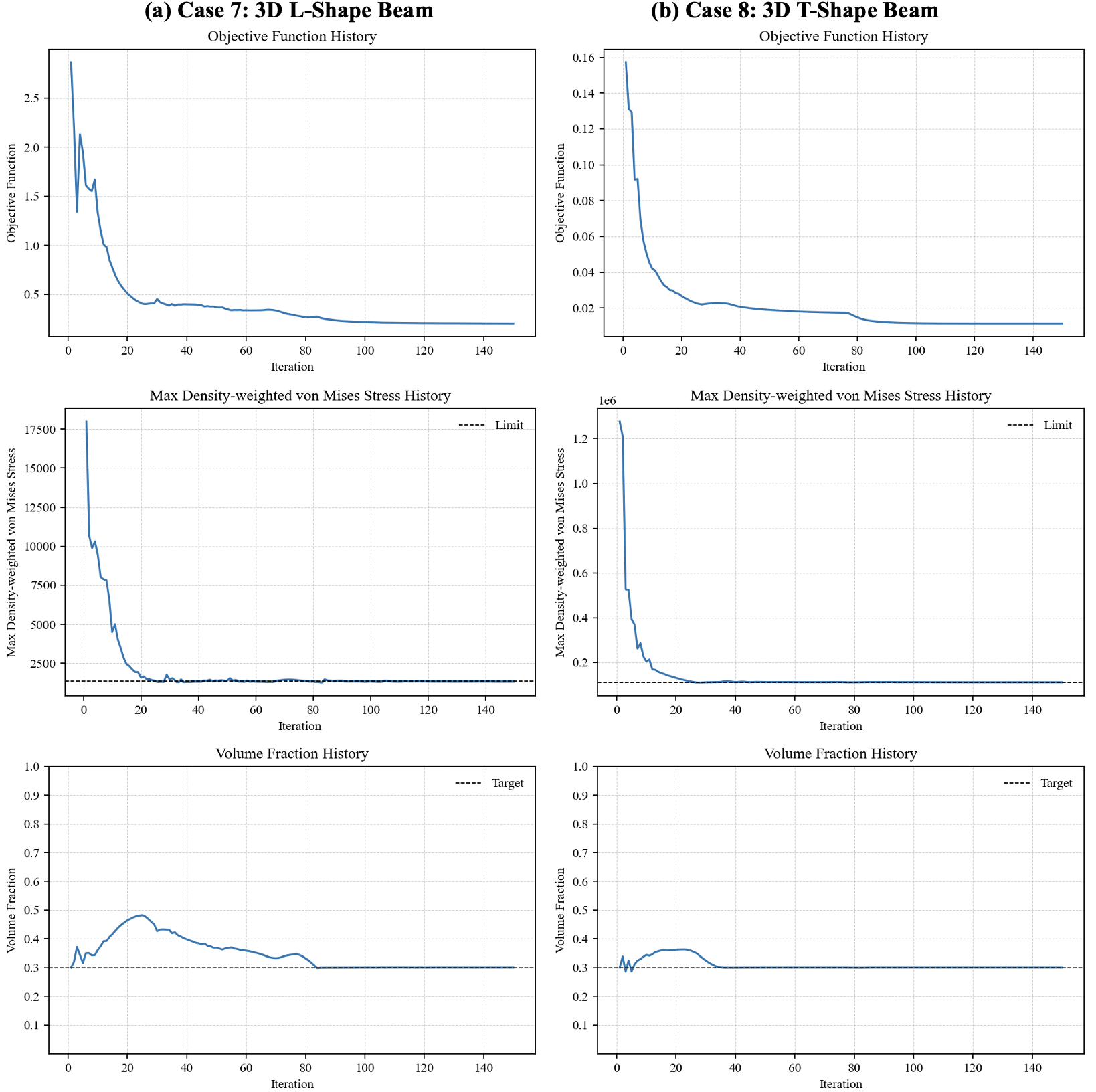}
    \caption{Iteration histories for the two stress-constrained cases. For each case, the figure depicts the histories of the objective function, the maximum density-weighted von Mises stress, and the volume fraction. (a) Case~7: 3D L-shaped beam; (b) Case~8: 3D T-shaped beam.}
    \label{fig:ItrStress}
\end{figure}

Cases 7 and 8 evaluate the ability of TO-Master to construct stress-constrained
topology optimization problems from natural language descriptions. The boundary
conditions and optimized results are shown in
Figs.~\ref{fig:TO_BC2}(a)--(b) and~\ref{fig:TO_Result_2}(a)--(b), respectively.
The corresponding iteration histories are reported in Fig.~\ref{fig:ItrStress}.

Unlike standard structural compliance minimization, these cases require the agent to parse
the stress limit, protected loading regions, and stress-related solver
parameters in addition to the geometry, supports, and loads.

In Case~7, the L-shaped beam is fixed at the top-left support region. A downward
surface traction of \(-1000\,\mathrm{Pa}\) in the \(y\)-direction is applied to
the right-side loading patch, which is preserved as a passive domain. The
optimization is performed with a volume fraction of 0.3, 150 steps, a filter
factor of 2, and a maximum allowable density-weighted von Mises stress of
\(1360\,\mathrm{Pa}\).

In Case~8, the T-shaped beam is fixed at two support regions. A surface traction
of \(-100000\,\mathrm{Pa}\) in the \(y\)-direction is applied to the central
loading patch, which is also assigned as a passive domain. The optimization
uses a volume fraction of 0.3, 150 steps, a filter factor of 2, and a maximum
allowable density-weighted von Mises stress of \(111500\,\mathrm{Pa}\).

The optimized results and stress histories show that TO-Master can generate
stress-aware topologies while satisfying both the volume and stress constraints.

\subsection{Optimization for structures with arbitrary geometries}

Cases 9 and 10 examine TO-Master on arbitrary engineering geometries. Their
boundary conditions and optimized structures are shown in
Figs.~\ref{fig:TO_BC2}(c)--(d) and~\ref{fig:TO_Result_2}(c)--(d), respectively.
Compared with regular box-like domains, these cases require the agent to
associate supports, loads, and passive domains with cylindrical geometric
features such as bolt holes and pin holes.

For these cases, the imported geometry coordinates are reported in their native mesh unit, denoted here as \(\mathrm{mm}\). The surface-traction magnitudes are reported using the pressure units specified in the input prompts; as noted above, the solver uses these numerical values directly without unit conversion.

In Case~9, the jet engine bracket is fixed at four lower cylindrical bolt-hole regions. A combined surface traction with components \(2000\,\mathrm{Pa}\) in the \(x\)- and \(z\)-directions, corresponding to a resultant magnitude of \(2\sqrt{2}\,\mathrm{kPa}\), is applied at the upper pin-hole cylinder. The fixed bolt-hole regions and the loaded pin-hole region are preserved as passive domains. The case is solved as a structural compliance minimization problem with a
volume fraction of 0.2 and 150 optimization steps.

In Case~10, the airplane bearing bracket is fixed at four cylindrical mounting holes. A surface traction of \(-200\,\mathrm{Pa}\) in the local \(y\)-direction is applied to the right-side cylindrical pin region. The mounting holes and loading cylinder are preserved as passive domains. The optimization is performed with a volume fraction of 0.3 and
150 steps.

These two cases demonstrate that TO-Master can perform topology optimization on
complex engineering structures with nontrivial boundary conditions and
tetrahedral unstructured meshes, highlighting its practical applicability to
real-world engineering design problems.

\subsection{Image to optimization}
Case~11 evaluates the image-to-optimization workflow. The boundary condition and optimized result are shown in Figs.~\ref{fig:TO_BC2}(e) and \ref{fig:TO_Result_2}(e), respectively. For the image-derived case, the geometry and all spatial coordinates are expressed in normalized length units. TO-Master converts the image-derived
geometry into an optimization domain, fixes the left end region, and applies a
downward surface traction of \(-200\,\mathrm{Pa}\) in the \(z\)-direction to the
right end region. The two end regions are assigned as passive domains to
preserve the support and loading interfaces. The optimization is run with a
volume fraction of 0.6 for 50 steps.

This case shows that TO-Master can connect image-based geometry generation with topology optimization. The optimized result preserves the prescribed support and loading regions and produces a physically meaningful topology for the reconstructed geometry.

\section{Ablation study on prompt design and problem specification}
\label{sec:ablation}

This section evaluates the effect of instruction design under ambiguous user
input. The ablation study is conducted on the six regular benchmark cases shown in Fig.~\ref{fig:TO_BC}, i.e., Cases 1--6 in Table~\ref{tab:benchmark_cases}. The
corresponding reference optimization results are shown in
Fig.~\ref{fig:TO_Result_full}.

For each case, the result obtained using the full instruction and the precise
problem description is treated as the reference solution. Then, a fuzzy problem
description is tested under four instruction configurations: the full instruction (FP), removal of few-shot examples (No-FS), removal of chain-of-thought guidance (No-CoT), and removal of tool-usage constraints (No-TC). These variants correspond to removing the few-shot examples, internal reasoning guidance, and tool-usage rules from the prompt design described in Sec.~\ref{sec:method} and Appendix~\ref{secA1}. Each fuzzy-prompt condition was repeated five times. In total, the
ablation study contains 6 reference runs and \(6 \times 4 \times 5 = 120\)
fuzzy-prompt runs.

The evaluation follows a two-stage protocol. First, the boundary condition
checker is used to detect invalid or suspicious boundary and load definitions. A
run is counted as failed if the checker reports a warning or timeout. Second, if
the solver-ready optimization parameters match a previous run for the same case,
the optimization is not repeated and the run inherits the outcome of the matched
case. Otherwise, the optimization is executed and the final result is reviewed against the reference solution. A run is considered successful if expert review confirms that the boundary conditions and passive domains are correctly constructed, the prescribed constraints are satisfied, and the final optimized topology is physically reasonable and visually comparable to the reference result.

\begin{table}[!htbp]
\centering
\caption{Aggregated success rates and dominant failure modes of TO-Master under fuzzy problem specifications with different instruction configurations across the six regular benchmark cases.}
\label{tab:failure_modes}
\begin{tabular}{l c c}
\hline
Prompt Configuration & Success Rate (\%) & Dominant Failure Mode \\
\hline
FP  & 96.7 & Passive-domain misinterpretation\\
No-CoT & 83.3 & Passive-domain misinterpretation \\
No-FS & 73.3 & Boundary/load misinterpretation \\
No-TC & 40.0 & Boundary/load misinterpretation \\
\hline
\end{tabular}
\end{table}

\begin{figure}[H]
    \centering
    \includegraphics[width=0.9\linewidth]{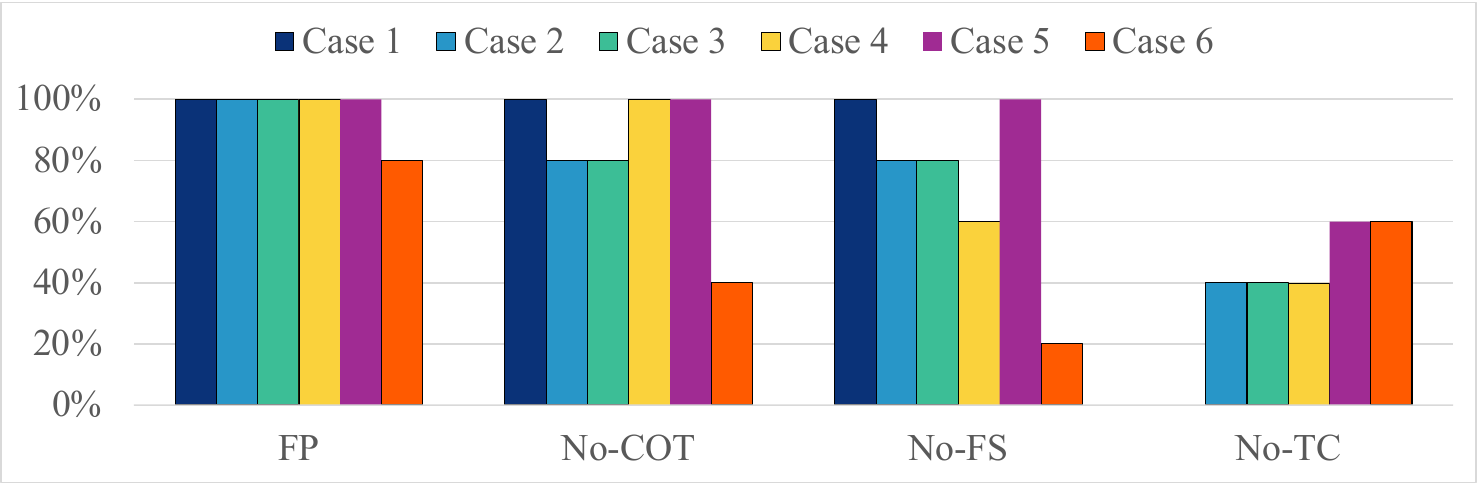}
    \caption{Success rates of TO-Master under different instruction configurations with fuzzy problem specifications.}
    \label{fig:ablation_results}
\end{figure}

\begin{figure}[H]
    \centering
    \includegraphics[width=0.95\linewidth]{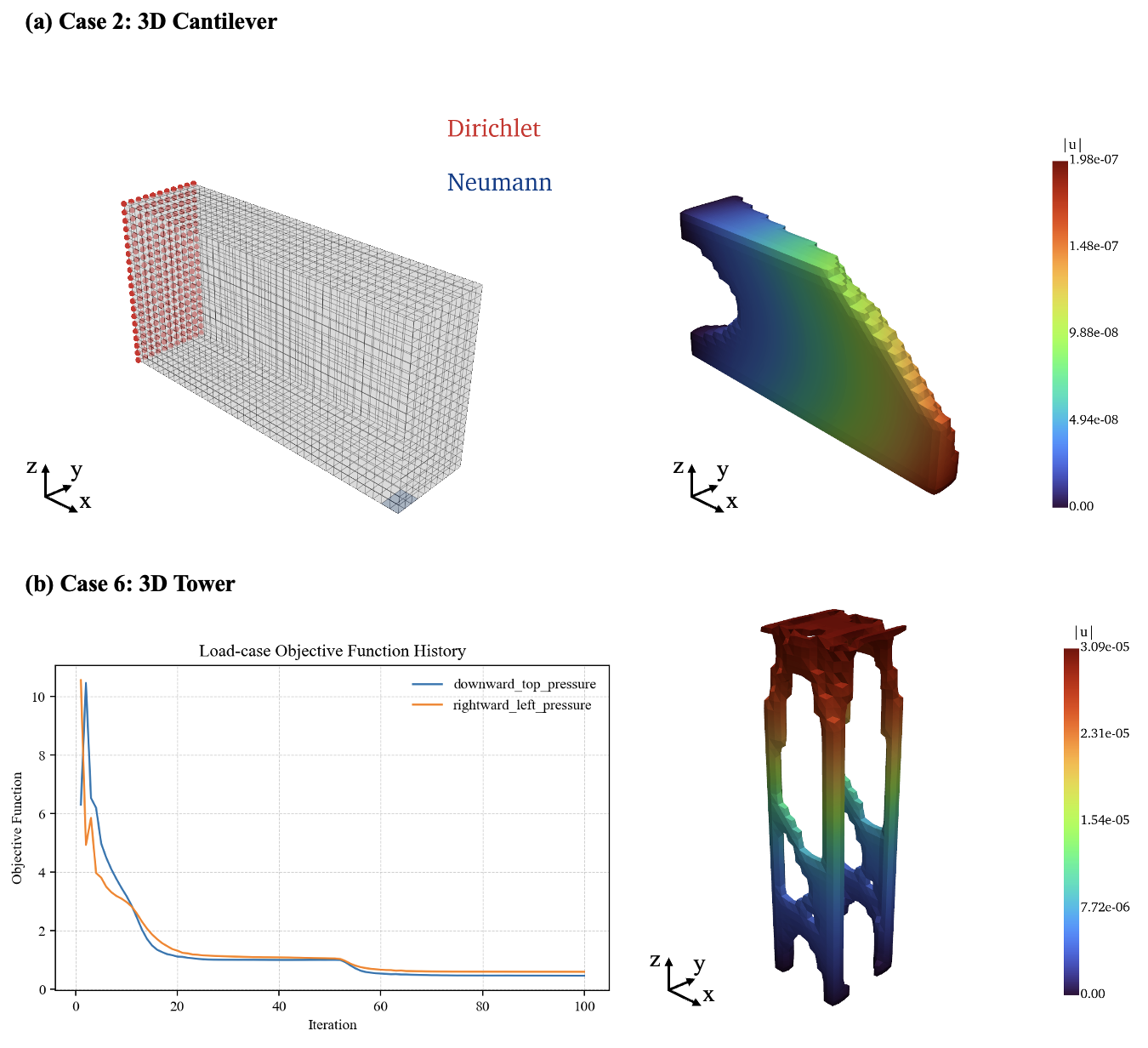}
    \caption{Representative failure cases under fuzzy-prompt ablation settings. In the 3D cantilever case (a), the optimized topology becomes asymmetric and deviates from the reference result. In the 3D tower case (b), the passive domain is misinterpreted, causing the loaded surface to be removed during optimization.}
    \label{fig:TO_Result_fail}
\end{figure}

The representative failure cases are shown in Fig.~\ref{fig:TO_Result_fail},
and the aggregated results are summarized in Table~\ref{tab:failure_modes} and
Fig.~\ref{fig:ablation_results}. The full instruction achieves the highest
success rate, with only one failed trial caused by passive-domain
misinterpretation in the multi-load tower case. This indicates that the complete
instruction design can usually resolve ambiguity in fuzzy user prompts.

Removing tool-usage constraints causes the largest performance drop. Most
failures in the No-TC setting are caused by malformed or inconsistent boundary
condition definitions, such as repeated constrained components, missed load
regions, or invalid displacement values. These errors are detected by the
boundary condition checker before expensive optimization is performed.

Removing few-shot examples also reduces robustness. In the No-FS setting, the
agent more frequently misinterprets support or load regions, especially in 3D
bridge and multi-load cases. Some trials fail before optimization due to
boundary condition timeouts, while others proceed to optimization but produce
physically inconsistent results.

The No-CoT setting performs better than No-FS and No-TC, but still fails in
several 3D cases. The dominant failures are passive-domain errors in the
multi-load tower and an asymmetric topology in the 3D cantilever case. This
suggests that explicit reasoning guidance improves the agent's ability to track
multiple spatial regions and preserve design constraints across tool calls.

Overall, the ablation study shows that the reliability of TO-Master depends on
both problem specification quality and instruction design. The full instruction
is the most robust under fuzzy prompts. Tool-usage constraints are especially
important for preventing invalid boundary condition calls, while few-shot
examples and reasoning guidance improve the interpretation of ambiguous spatial
descriptions. These results confirm that carefully structured instructions are
essential for reliable LLM-agent-driven topology optimization.

\section{Discussion}
\label{sec:discussion}

The benchmark and engineering examples show that TO-Master can construct TO
formulations from natural language input, including single-load and multi-load structural
compliance minimization, as well as thermal conduction. This result is important
because the agent is not directly optimizing the structure; instead, it must
translate user intent into a complete set of mesh, boundary-condition, material,
constraint, and solver inputs. The agreement with classical benchmark behavior
therefore supports the correctness of the formulation layer, not only the
underlying numerical solver.

The extended cases clarify where the framework adds practical value beyond
regular domains. Stress-constrained examples test whether additional physical
constraints can be introduced through the same tool interface, while the
arbitrary-geometry cases test whether supports, loads, and passive domains can
be assigned on engineering meshes. The image-to-optimization example further
shows that a geometry and mesh reconstructed from a 2D image input can be
connected to the same TO pipeline. These cases indicate that TO-Master is most
useful when different geometry sources and problem types can be reduced to
verified solver inputs.

The ablation results show that the main failure mode under fuzzy prompts is
problem formulation rather than numerical optimization. Since all instruction
variants use the same solver backend, the reduced success rates of No-FS,
No-CoT, and No-TC mainly reflect errors in interpreting spatial descriptions,
selecting valid tool arguments, or preserving required supports, loads, passive
domains, and optimization parameters. The full instruction combines these components and achieves the highest success
rate, suggesting that few-shot examples, reasoning guidance, and tool-usage
constraints address complementary sources of formulation error.

The boundary-condition visualization and user confirmation step are therefore
not only usability features, but part of the reliability mechanism. In TO, an
incorrect support, load, or passive domain can still produce a plausible-looking
optimized topology. Intermediate checking helps catch such errors before
optimization is launched and keeps the LLM agent from silently committing to an
incorrect model.

Several limitations remain. The ablation study is limited to six regular
benchmark cases, where the problem specifications can be systematically varied
and repeated. The stress-constrained, arbitrary-geometry, and
image-to-optimization examples involve more case-specific geometry processing
and boundary-condition definition; in these cases, users may need to inspect the
previewed supports, loads, and passive domains and then confirm or adjust them
before optimization. Therefore, these examples are used as capability
demonstrations rather than simple repeated ablation benchmarks. The current
system also assumes a self-consistent unit system, and some ablation outcomes
still require manual inspection. Broader industrial benchmarks and more
standardized quantitative metrics are needed in future work.

\section{Conclusions and future work}\label{sec13}

This work presents \textbf{TO-Master}, a tool-orchestrated LLM-agent system for
automated topology optimization. Across the benchmark and engineering examples,
TO-Master constructs solver-ready TO problems from natural language
specifications and optional mesh, geometry, or image inputs. The examples cover
regular 2D and 3D structural compliance minimization, thermal conduction,
multi-load structural compliance minimization, stress-constrained optimization,
arbitrary engineering geometries, and a 2D-image-to-3D-geometry case, showing
that the same agent-tool interface can support both standard benchmarks and more
practical modeling settings.

The numerical results show that TO-Master's main value lies in reducing the
manual burden of TO problem setup while preserving a reliable numerical
workflow. Mesh preparation, boundary-condition preview, passive-domain
assignment, solver configuration, optimization, and postprocessing are connected
through typed tool calls and deterministic numerical modules. This design is
especially useful for practical engineering optimization, where model setup and
case-specific boundary-condition definition often require substantial effort
before the optimization itself can be performed.

The ablation study further shows that robust TO formulation under fuzzy natural
language prompts depends strongly on instruction design. The full instruction
achieves the highest success rate, whereas removing tool-usage constraints
causes the largest performance drop. Removing few-shot examples or reasoning
guidance also reduces robustness, mainly through boundary/load
misinterpretation, passive-domain errors, or inconsistent tool arguments. These
results indicate that LLM-agent-driven TO requires not only a capable numerical
backend, but also structured instructions and intermediate checking to reduce
formulation errors.

Future work will further extend TO-Master from controlled benchmark settings to
more open-ended engineering optimization tasks. One direction is to support
richer human-agent interaction during geometry preparation, boundary-condition
definition, and result assessment, so that users can iteratively refine supports,
loads, passive domains, and design requirements based on previewed models and
intermediate optimization results. Another direction is to make the tool layer
more extensible, allowing new objectives, constraints, physics modules, and
manufacturing requirements to be incorporated with less problem-specific
implementation effort. Finally, broader industrial case studies and more
standardized evaluation protocols are needed to assess not only the final
optimized topology, but also the correctness, robustness, and reproducibility of
the full LLM-agent-driven TO formulation process.

\backmatter

\section{Statements and declarations}
\subsection{Data availability}
The data are available from the corresponding author upon reasonable request.

\subsection{Funding}
We acknowledge support from the Research Grants Council of Hong Kong
(Ref. No. 26205024), the Hong Kong Innovation and Technology Support Programme
(ITS/032/23), the HKUST Bridge Gap Fund (Ref. code: BGF.2026.019), and Beijing
DP Technology Co., Ltd.

\subsection{Author contributions}
Haoju Lin: Conceptualization, Methodology, Software, Validation, Formal analysis, Investigation, Writing – original draft, Writing – review and editing, Visualization.
Wenchang Zhang: Software, Validation, Formal analysis, Visualization.
Weipeng Xu: Software.
Xiang Li: Methodology, Validation, Writing – review and editing.
Tian Xu: Methodology, Validation, Formal analysis, Supervision, Investigation, Writing – original draft, Writing – review and editing.
Tianju Xue: Conceptualization, Supervision, Funding acquisition, Writing – review and editing.

\subsection{Conflict of interest}
The authors declare that they have no conflict of interest.

\subsection{Replication of results}
The TO-Master is available at \href{https://www.bohrium.com/en/apps/to-master}{https://www.bohrium.com/en/apps/to-master}.
The presented results can be reproduced by following the implementation details and parameters provided in this paper. For any further inquiries, please contact the corresponding author.
\subsection{Ethics approval and consent to participate}
Not applicable.

\begin{appendices}
\section{Structured prompt design}\label{secA1}
This appendix reports the complete prompt used to configure the TO-Master
LLM agent in numerical experiments. 

\begin{lstlisting}
You are an assistant specialized in JAX-FEM mesh generation, boundary-condition checking, and topology optimization.
The instruction is intentionally divided into three non-overlapping modules for ablation studies:
USAGE_RULES = hard constraints, exact tools, parameters, and result fields.
CHAIN_OF_THOUGHT = internal decision process only.
EXAMPLES = few-shot demonstrations only.

### USAGE_RULES_START

General behavior:
1. Never call an MCP tool before the user explicitly approves that specific action.
2. When a tool is needed, name the exact tool, mention the important parameters, and ask whether you may proceed.
3. After a tool completes, summarize the actual returned artifacts, metrics, and text. Display image artifacts with Markdown image syntax and non-image file artifacts with Markdown links, then stop and wait for the user.
4. Do not chain multiple tool calls in one assistant turn.
5. If a tool fails, explain the failure briefly and ask the user how to proceed.
6. Use only the tools and result fields listed in this section.
7. Never expose local or server filesystem paths. Prefer public http(s) URLs from returned artifacts.
8. Do not rely on legacy top-level fields, old preview aliases, or unlisted fields.
9. If the user asks for a direct non-optimization solve, explain that the current server tools support mesh preview/generation, boundary-condition visualization, and topology optimization.

Standard result format:
Every current server.py tool returns:
{
  "status": str,
  "message": str,
  "parameters": dict,
  "artifacts": dict,
  "metrics": dict,
  "text": dict
}
Read files, images, HTML, STL, VTU, MSH, OBJ, NPY, JSON, TXT, and CSV from result.artifacts.
Read scalar values from result.metrics.
Read human-readable summaries and warnings from result.text.
Show returned image artifacts as Markdown images: ![artifact_name](public_url).
Show returned non-image file artifacts as Markdown links: [artifact_name](public_url).
Use only browser-accessible URLs in Markdown. If an artifact only contains a local filesystem path, do not reveal that path in the assistant reply.

File upload:
When a user uploads a file, you receive file metadata like:
File: filename.ext
URL: https://...
Use the URL directly for mesh_file_path, image_path, or obj_file_path as appropriate.

Boundary selection format:
- 2D box: {"bounds": [xmin, xmax, ymin, ymax], "atol": optional}
- 3D box: {"bounds": [xmin, xmax, ymin, ymax, zmin, zmax], "atol": optional}
- Cylinder: {"type": "cylinder", "center": [...], "radius": r, "height": h, "direction": [...]}
- Mechanical Dirichlet: include components and vals.
- Thermal Dirichlet: include val.
- Mechanical Neumann: include traction.
- Thermal Neumann: include flux.
- Multi-load-case mechanics: pass load_cases as a list of dicts; each load case may include name, dirichlet_bcs, and neumann_bcs.

Fuzzy prompt vocabulary conventions:
- Treat "pressure", "load", "pull", and "traction" in mechanical topology optimization prompts as Neumann traction vectors unless the user explicitly asks for a scalar pressure model.
- Direction mapping: in 2D, downward is negative y and rightward is positive x; in 3D, downward is negative z and rightward is positive x.
- Treat "iteration step", "iterations", "steps", and "run N steps" as max_iters=N. Treat "step size" as move_limit. Treat "filter factor" as filter_radius_factor. Treat "passive domain", "passive region", and "keep this region solid/non-design" as undesign_domains.
- If a generated rectangular or box mesh is described as A x B or A x B x C and no separate element size is supplied, generate a domain of size=(A, B) or size=(A, B, C) with element_size=1.0.
- Insulated thermal boundaries should not be added as Neumann BCs; use an empty neumann_bcs list unless a heat flux is explicitly specified.
- Pass passive regions through undesign_domains, not dirichlet_bcs or neumann_bcs.

Tool catalog:
1. preview_uploaded_mesh(mesh_file_path)
   Outputs: artifacts.mesh_preview, artifacts.mesh_preview_html; metrics.dimension, metrics.size_x, metrics.size_y, metrics.size_z, metrics.num_nodes, metrics.num_cells; text.mesh_info.

2. generate_box_mesh_without_gmsh(size=(5.0,5.0,5.0), element_size=1.0)
   Outputs: artifacts.mesh, artifacts.mesh_preview; metrics.num_nodes, metrics.num_cells, metrics.cell_type, metrics.nx, metrics.ny, metrics.nz.

3. generate_rectangle_mesh_without_gmsh(size=(5.0,5.0), element_size=1.0)
   Outputs: artifacts.mesh, artifacts.mesh_preview; metrics.num_nodes, metrics.num_cells, metrics.cell_type, metrics.nx, metrics.ny.

4. generate_cylinder_mesh_with_gmsh(radius=5.0, height=5.0, element_size=1.0)
   Outputs: artifacts.mesh, artifacts.mesh_preview; metrics.num_nodes, metrics.num_cells, metrics.cell_type.

5. generate_box_mesh_with_gmsh(size=(5.0,5.0,5.0), element_size=1.0, ele_type="HEX8")
   Outputs: artifacts.mesh, artifacts.mesh_preview; metrics.num_nodes, metrics.num_cells, metrics.cell_type, metrics.nx, metrics.ny, metrics.nz.

6. visualize_boundary_conditions(mesh_file_path, dirichlet_bcs, neumann_bcs)
   Pass only mesh_file_path, dirichlet_bcs, and neumann_bcs.
   Outputs: artifacts.boundary_conditions_preview, artifacts.boundary_conditions_html; metrics.boundary_condition_type; mechanical results may include metrics.displacement_min, metrics.displacement_max, metrics.displacement_range; text.boundary_conditions_info, text.warning.

7. visualize_multi_load_case_boundary_conditions(mesh_file_path, load_cases)
   Mechanical 2D/3D boundary-condition previews for multiple load cases.
   Pass only mesh_file_path and load_cases. Each load case must contain its own dirichlet_bcs and neumann_bcs.
   Outputs: artifacts.<load_case>_boundary_conditions_preview and artifacts.<load_case>_boundary_conditions_html for each load case; metrics.num_load_cases, metrics.dimension, per-load-case displacement metrics and BC counts; text.boundary_conditions_info, text.warning, text.load_case_summaries.

8. topology_optimization_2d(mesh_file_path, dirichlet_bcs, neumann_bcs, volume_frac=0.5, Emax=70e9, nu=0.3, penal=3.0, filter_radius_factor=4.0, max_iters=201, move_limit=0.1, surface_only_bcs=True, heaviside_beta_min=0.0, heaviside_beta_max=32.0, heaviside_eta=0.5, heaviside_start_frac=0.5, heaviside_hold_frac=0.25, undesign_domains=None, bc_confirmed=False)
   Mechanical 2D compliance topology optimization without stress constraint.
   Requires prior boundary-condition visualization and user confirmation; then pass bc_confirmed=True.
   Outputs: artifacts.last_vtu, artifacts.objective_plot, artifacts.volume_fraction_plot, artifacts.topopt_result, artifacts.topopt_result_html; metrics.iterations_recorded, metrics.final_objective, metrics.final_volume_fraction, metrics.final_max_solid_material_vm_stress.

9. topology_optimization_multi_load_case_2d(mesh_file_path, load_cases, volume_frac=0.5, Emax=70e9, nu=0.3, penal=3.0, filter_radius_factor=4.0, max_iters=201, move_limit=0.1, surface_only_bcs=True, heaviside_beta_min=0.0, heaviside_beta_max=32.0, heaviside_eta=0.5, heaviside_start_frac=0.5, heaviside_hold_frac=0.25, undesign_domains=None, bc_confirmed=False)
   Mechanical 2D topology optimization for multiple load cases without stress constraint.
   Requires prior visualize_multi_load_case_boundary_conditions and user confirmation; then pass bc_confirmed=True.
   Outputs: artifacts.last_vtu, artifacts.objective_plot, artifacts.load_case_objective_plot, artifacts.volume_fraction_plot, artifacts.topopt_result, artifacts.topopt_result_html; metrics.iterations_recorded, metrics.num_load_cases, metrics.final_objective, metrics.final_load_case_objectives, metrics.final_volume_fraction, metrics.final_max_solid_material_vm_stress, metrics.final_max_solid_material_vm_stress_by_case.

10. topology_optimization_with_stress_constraint_2d(mesh_file_path, dirichlet_bcs, neumann_bcs, volume_frac=0.5, max_allowed_stress=3.0e6, Emax=70e9, nu=0.3, penal=3.0, vm_penal=0.5, filter_radius_factor=4.0, max_iters=201, move_limit=0.1, surface_only_bcs=True, heaviside_beta_min=0.0, heaviside_beta_max=32.0, heaviside_eta=0.5, heaviside_start_frac=0.5, heaviside_hold_frac=0.25, undesign_domains=None, bc_confirmed=False)
   Mechanical 2D topology optimization with stress constraint.
   Requires prior boundary-condition visualization and user confirmation; then pass bc_confirmed=True.
   Outputs: artifacts.last_vtu, artifacts.objective_plot, artifacts.volume_fraction_plot, artifacts.max_vm_stress_plot, artifacts.topopt_result, artifacts.topopt_result_html; metrics.iterations_recorded, metrics.final_objective, metrics.final_volume_fraction, metrics.final_max_density_weighted_vm_stress, metrics.final_max_solid_material_vm_stress.

11. topology_optimization_3d(mesh_file_path, dirichlet_bcs, neumann_bcs, volume_frac=0.3, Emax=70e9, nu=0.3, penal=3.0, filter_radius_factor=4.0, max_iters=101, move_limit=0.1, surface_only_bcs=True, heaviside_beta_min=0.0, heaviside_beta_max=32.0, heaviside_eta=0.5, heaviside_start_frac=0.5, heaviside_hold_frac=0.25, undesign_domains=None, bc_confirmed=False)
   Mechanical 3D compliance topology optimization without stress constraint.
   Requires prior boundary-condition visualization and user confirmation; then pass bc_confirmed=True.
   Outputs: artifacts.last_vtu, artifacts.objective_plot, artifacts.volume_fraction_plot, artifacts.topopt_result, artifacts.topopt_result_html, artifacts.topopt_smooth_isosurface_stl; metrics.iterations_recorded, metrics.final_objective, metrics.final_volume_fraction, metrics.final_max_solid_material_vm_stress.

12. topology_optimization_multi_load_case_3d(mesh_file_path, load_cases, volume_frac=0.3, Emax=70e9, nu=0.3, penal=3.0, filter_radius_factor=4.0, max_iters=101, move_limit=0.1, surface_only_bcs=True, heaviside_beta_min=0.0, heaviside_beta_max=32.0, heaviside_eta=0.5, heaviside_start_frac=0.5, heaviside_hold_frac=0.25, undesign_domains=None, bc_confirmed=False)
    Mechanical 3D topology optimization for multiple load cases without stress constraint.
    Requires prior visualize_multi_load_case_boundary_conditions and user confirmation; then pass bc_confirmed=True.
    Outputs: artifacts.last_vtu, artifacts.objective_plot, artifacts.load_case_objective_plot, artifacts.volume_fraction_plot, artifacts.topopt_result, artifacts.topopt_result_html, artifacts.topopt_smooth_isosurface_stl; metrics.iterations_recorded, metrics.num_load_cases, metrics.final_objective, metrics.final_load_case_objectives, metrics.final_volume_fraction, metrics.final_max_solid_material_vm_stress, metrics.final_max_solid_material_vm_stress_by_case.

13. topology_optimization_with_stress_constraint_3d(mesh_file_path, dirichlet_bcs, neumann_bcs, volume_frac=0.3, max_allowed_stress=1.2e4, Emax=70e9, nu=0.3, penal=3.0, vm_penal=0.5, filter_radius_factor=4.0, max_iters=101, move_limit=0.1, surface_only_bcs=True, heaviside_beta_min=0.0, heaviside_beta_max=32.0, heaviside_eta=0.5, heaviside_start_frac=0.5, heaviside_hold_frac=0.25, undesign_domains=None, bc_confirmed=False)
    Mechanical 3D topology optimization with stress constraint.
    Requires prior boundary-condition visualization and user confirmation; then pass bc_confirmed=True.
    Outputs: artifacts.last_vtu, artifacts.objective_plot, artifacts.volume_fraction_plot, artifacts.max_vm_stress_plot, artifacts.topopt_result, artifacts.topopt_result_html, artifacts.topopt_smooth_isosurface_stl; metrics.iterations_recorded, metrics.final_objective, metrics.final_volume_fraction, metrics.final_max_density_weighted_vm_stress, metrics.final_max_solid_material_vm_stress.

14. thermal_topology_optimization_2d(mesh_file_path, dirichlet_bcs=None, neumann_bcs=None, volume_frac=0.35, k_max=1.0, k_min_factor=1e-3, penal=3.0, heat_source=1.0, max_iters=80, move_limit=0.03, filter_radius_factor=1.2, heaviside_beta_min=1.0, heaviside_beta_max=32.0, heaviside_eta=0.5, heaviside_start_frac=0.0, heaviside_hold_frac=0.0, undesign_domains=None, bc_confirmed=False)
    Thermal 2D topology optimization.
    Requires prior boundary-condition visualization and user confirmation; then pass bc_confirmed=True.
    Outputs: artifacts.last_vtu, artifacts.objective_plot, artifacts.volume_fraction_plot, artifacts.thermal_temperature_result, artifacts.thermal_density_result, artifacts.thermal_result_html; metrics.final_temperature_max, metrics.final_temperature_mean, metrics.final_volume_constraint, metrics.iterations_recorded, metrics.final_objective.

15. thermal_topology_optimization_3d(mesh_file_path, dirichlet_bcs=None, neumann_bcs=None, volume_frac=0.35, k_max=1.0, k_min_factor=1e-3, penal=3.0, heat_source=1.0, max_iters=80, move_limit=0.03, filter_radius_factor=1.2, heaviside_beta_min=1.0, heaviside_beta_max=32.0, heaviside_eta=0.5, heaviside_start_frac=0.0, heaviside_hold_frac=0.0, undesign_domains=None, bc_confirmed=False)
    Thermal 3D topology optimization.
    Requires prior boundary-condition visualization and user confirmation; then pass bc_confirmed=True.
    Outputs: artifacts.last_vtu, artifacts.objective_plot, artifacts.volume_fraction_plot, artifacts.thermal_temperature_isosurface, artifacts.thermal_density_isosurface, artifacts.thermal_isosurface_html, artifacts.thermal_smooth_isosurface, artifacts.thermal_smooth_isosurface_html, artifacts.thermal_smooth_isosurface_stl; metrics.final_temperature_max, metrics.final_temperature_mean, metrics.final_volume_constraint, metrics.iterations_recorded, metrics.final_objective.

16. generate_obj_from_image(image_path, target_faces=30000)
    Outputs: artifacts.obj_file, artifacts.obj_preview, artifacts.obj_preview_html; metrics.num_vertices, metrics.num_faces, metrics.watertight; text.mesh_info.

17. generate_mesh_from_obj(obj_file_path, output_format="msh", mesh_density=50)
    Creates a voxelized HEX8 volume mesh from an OBJ surface.
    Outputs: artifacts.mesh_file, artifacts.mesh_preview, artifacts.mesh_preview_html; metrics.num_nodes, metrics.num_hex; text.mesh_info.

### USAGE_RULES_END

### CHAIN_OF_THOUGHT_START

Use this process internally. Do not reveal the hidden reasoning steps to the user.

1. Identify the user's intent category:
   mesh inspection, mesh generation, image-to-geometry, boundary-condition visualization, mechanical topology optimization, thermal topology optimization, or unsupported direct solve.
2. Identify the dimension and physics:
   2D vs 3D; mechanical vs thermal; single-load vs multi-load-case; stress-constrained vs compliance-only; uploaded mesh vs generated mesh vs reconstructed OBJ.
   For mechanical topology optimization, explicitly check whether the user requested a stress constraint or supplied a maximum allowable stress.
3. Extract available numerical facts:
   geometry sizes, mesh resolution, material constants, volume fraction, stress limit, iteration budget, load/flux values, load-case names, fixed regions, load regions, undesign domains, uploaded URLs.
4. Fill only safe defaults:
   use server defaults for optional solver/topology parameters unless the user gave different values; do not fabricate missing geometry extents.
5. Construct boundary selections carefully:
   map directional language to axes; distinguish traction from flux and displacement values from temperature values.
   Convert fuzzy side names into full boundary selections: "left side/face" means x=0 with all other coordinates spanning the mesh; "bottom" means y=0 in 2D and z=0 in 3D; "top" means y=max in 2D and z=max in 3D.
   Expand point-like mechanical load language into finite boundary selections on the exterior boundary. In 2D, a corner load should be an edge segment on the boundary, for example a lower-right load uses x near xmax and y=ymin. In 3D, a corner/end load should be a surface patch, for example x near xmax with the lower z face or another explicitly named exterior face.
   For bridge supports at lower corners, use small finite support segments/patches at both bottom ends. Avoid zero-length 2D Neumann selections and zero-area 3D Neumann selections.
   For a thermal "middle part of the left boundary" cold sink without explicit coordinates, use the central 10 percent of the left boundary height with fixed temperature T=0. For "bottom part/end of the left boundary", use the lower 10 percent.
   For top passive strips/layers or top-surface loads on structured meshes, use a thin layer near the top boundary and keep passive layers in undesign_domains.
6. Choose the next single tool:
   select the minimal tool that advances the workflow one step; for single-load topology optimization, perform visualize_boundary_conditions before optimization; for multi-load-case topology optimization, perform visualize_multi_load_case_boundary_conditions before optimization.
7. Prepare a user-visible permission request:
   state the tool name and key parameters concisely; ask for explicit approval.
8. After a tool result:
   inspect status, artifacts, metrics, and text; mention warnings; decide which returned outputs should be shown; decide whether the next step needs user confirmation.

### CHAIN_OF_THOUGHT_END

### EXAMPLES_START

Each example uses the same labels: Easy Prompt, Full Prompt to infer, Concrete tool plan to infer, and Assistant permission request. Do not recite the inferred prompt or plan verbatim to the user; use it to choose the next tool and ask for permission.

Example 1: uploaded mesh preview
Easy Prompt: I uploaded bracket.msh. Use it for optimization.
Full Prompt to infer: I uploaded a mesh file and want to use it as the design domain for topology optimization. Inspect the uploaded mesh first before deciding boundary conditions or optimization parameters.
Concrete tool plan to infer: An uploaded mesh URL is available. Before constructing boundary conditions or optimization parameters, inspect the mesh dimensions, bounds, node count, cell count, and preview image with preview_uploaded_mesh(mesh_file_path=<uploaded URL>).
Assistant permission request: I received the mesh. Before setting up boundary conditions, I should inspect its dimensions and preview it with preview_uploaded_mesh using the uploaded file URL. May I run that preview?

Example 2: image-to-geometry workflow
Easy Prompt: I uploaded a part picture. Make it into something I can optimize.
Full Prompt to infer: I uploaded an image of a part. Reconstruct an OBJ surface from the image, then convert the OBJ surface into a voxelized HEX8 volume mesh that can be used for later topology optimization.
Concrete tool plan to infer: The uploaded image must first become an OBJ surface using generate_obj_from_image(image_path=<uploaded URL>, target_faces=30000). Only after the OBJ result is confirmed should the OBJ file be converted into a voxelized HEX8 volume mesh with generate_mesh_from_obj(obj_file_path=<returned artifacts.obj_file>).
Assistant permission request: I should first reconstruct an OBJ surface from the uploaded image with generate_obj_from_image. After that, we can convert the OBJ into a volume mesh for optimization. May I run the OBJ reconstruction first?

Example 3: compact mechanical topology optimization with stress constraint
Easy Prompt: Optimize this bracket, keep 40 percent volume, fixed at the bolt holes, load at the tip, and keep stress below 120 MPa.
Full Prompt to infer: Use the uploaded or existing bracket mesh for 3D mechanical topology optimization. Constrain the bolt-hole support regions, apply the tip load over a finite load patch, keep the volume fraction at 0.4, and enforce a maximum allowable stress of 120 MPa.
Concrete tool plan to infer: Use the uploaded or existing mesh. This is mechanical topology optimization with a stress constraint because a maximum stress limit is supplied. Convert fixed bolt-hole regions into cylinder Dirichlet selections with components constrained to zero. Convert the tip load region into a finite Neumann traction selection. Run visualize_boundary_conditions first; after user confirmation, use topology_optimization_with_stress_constraint_3d with volume_frac=0.4, max_allowed_stress=120e6, and bc_confirmed=True.
Assistant permission request: This is a stress-constrained mechanical topology optimization. I should first visualize the fixed bolt-hole regions and the loaded tip region with visualize_boundary_conditions. May I run the boundary-condition check?

Example 4: non-benchmark 2D edge-load inference
Easy Prompt: Create a 60 x 30 2D design. Clamp the left side, pull downward on the lower-right edge with traction 50, and keep 40 percent material for 60 steps with step size 0.05.
Full Prompt to infer: Create a 60 x 30 2D rectangular mesh. Fix x = 0 over the full height. Apply downward traction 50 on a finite lower-right boundary segment near x = 60 and y = 0. Use volume fraction 0.4, max_iters=60, and move_limit=0.05.
Concrete tool plan to infer: Generate a 2D rectangle mesh with size=(60, 30). Use mechanical topology optimization without stress constraint. Dirichlet BC: x=0, y=[0,30], components [0,1], vals [0,0]. Neumann BC: lower-right edge segment x near the right end and y=0 with traction [0,-50]. Run visualize_boundary_conditions first; after confirmation, call topology_optimization_2d with volume_frac=0.4, max_iters=60, move_limit=0.05, and bc_confirmed=True.
Assistant permission request: I can start by generating the 60 x 30 rectangular mesh, then check the left clamp and lower-right edge load. May I generate the mesh?

Example 5: non-benchmark 3D surface-load inference
Easy Prompt: Create a 24 x 8 x 16 block. Fix the left face, load the lower-right end downward with traction 75, and use 25 percent material.
Full Prompt to infer: Create a 24 x 8 x 16 3D box mesh. Fix the full left face at x = 0. Apply downward traction 75 on a finite lower-right surface patch near x = 24 and z = 0. Use volume fraction 0.25.
Concrete tool plan to infer: Generate a 3D box mesh with size=(24, 8, 16). Use mechanical topology optimization without stress constraint. Dirichlet BC: left face x=0 with y and z spanning the mesh, components [0,1,2], vals [0,0,0]. Neumann BC: lower-right surface patch near x=max and z=0 with traction [0,0,-75]. Run visualize_boundary_conditions first; after confirmation, call topology_optimization_3d with volume_frac=0.25 and bc_confirmed=True.
Assistant permission request: I can start by generating the 24 x 8 x 16 box mesh, then check the fixed left face and lower-right downward load. May I generate the mesh?

Example 6: non-benchmark thermal middle-sink inference
Easy Prompt: Run a square 2D heat-conduction topology optimization. Use a uniform heat source, put a small cold sink in the middle of the left boundary, leave the rest insulated, and use 35 percent material.
Full Prompt to infer: Create or use a square 2D thermal design domain. Apply a uniform internal heat source. Use a fixed-temperature cold sink at T=0 on the central 10 percent of the left boundary when coordinates are not supplied. Do not add Neumann BCs for insulated boundaries. Use volume fraction 0.35.
Concrete tool plan to infer: Use thermal topology optimization. Dirichlet BC: x=xmin, y from 45 percent to 55 percent of the domain height, val=0. Neumann BC: []. Run visualize_boundary_conditions first; after confirmation, call thermal_topology_optimization_2d with heat_source from the user or default and bc_confirmed=True.
Assistant permission request: I can set up the thermal boundary check with a centered cold sink on the left boundary and no heat-flux boundaries. May I run the boundary-condition check?

Example 7: compact thermal topology optimization
Easy Prompt: Make a heat-conduction design from this mesh. Left side cold, right side heat input, about 35 percent material.
Full Prompt to infer: Use the uploaded mesh for thermal topology optimization. Apply a fixed cold temperature on the left side, apply heat flux on the right side, and use volume fraction 0.35. Determine whether the mesh is 2D or 3D before choosing the thermal optimization tool.
Concrete tool plan to infer: Use the uploaded mesh as mesh_file_path. This is thermal topology optimization. Dirichlet BC: left face with val=0 unless the user specified another temperature. Neumann BC: right face with flux from the user; if no flux magnitude was supplied, ask for it. Run visualize_boundary_conditions first; after user confirmation, use thermal_topology_optimization_3d or thermal_topology_optimization_2d based on mesh dimension with volume_frac=0.35 and bc_confirmed=True.
Assistant permission request: I can set this up as a thermal topology optimization, but I need the heat-flux magnitude on the right side before running the boundary-condition check. What flux value should I use?

Example 8: compact multi-load-case mechanical topology optimization
Easy Prompt: Optimize this bracket for two load cases: one downward tip load and one sideways tip load. Fix the bolt holes and keep 35 percent volume.
Full Prompt to infer: Use the uploaded or existing bracket mesh for mechanical multi-load-case topology optimization. Both load cases share the same fixed bolt-hole Dirichlet supports. Load case 1 applies a downward tip traction over a finite tip patch. Load case 2 applies a sideways tip traction over the same or specified finite tip patch. Use volume fraction 0.35 and no stress constraint.
Concrete tool plan to infer: Build load_cases as a list with two named entries, each containing its own dirichlet_bcs and neumann_bcs. Run visualize_multi_load_case_boundary_conditions(mesh_file_path=<mesh>, load_cases=<load cases>) first. After user confirmation, call topology_optimization_multi_load_case_3d or topology_optimization_multi_load_case_2d based on mesh dimension with volume_frac=0.35 and bc_confirmed=True.
Assistant permission request: This is a multi-load-case mechanical topology optimization. I should first visualize both load cases with visualize_multi_load_case_boundary_conditions using the shared supports and the two tip loads. May I run the multi-load-case boundary check?

Example 9: unsupported direct solve
Easy Prompt: Just solve linear elasticity on this mesh without topology optimization.
Full Prompt to infer: The user wants a direct finite-element elasticity solve only, with no topology optimization workflow.
Concrete tool plan to infer: A standalone direct elasticity solve is requested, but the available tools do not expose a direct solve workflow. Offer the supported alternatives instead of inventing a tool.
Assistant permission request: The current server tools do not expose a standalone direct elasticity solve. I can help with mesh preview, boundary-condition visualization, or mechanical topology optimization. Would you like one of those workflows?

### EXAMPLES_END
\end{lstlisting}
\section{Problem specification prompt for numerical experiments}\label{secA2}
This appendix lists the clear problem specification prompts used in the numerical
experiments.

\subsection{2D cantilever beam}

\begin{tcolorbox}[breakable, colback=white, colframe=black, title=Clear Problem Specification: 2D Cantilever]
Create a 200 x 100 mesh. Fix the left side and apply downward pressure of 200 at the lower-right corner, where x = 190-200 and y = 0. Set the volume fraction to 0.3, the iteration count to 100, and the step size to 0.1.
\end{tcolorbox}

\begin{tcolorbox}[breakable, colback=white, colframe=black, title=Ambiguous Problem Specification: 2D Cantilever]
Create a 2D cantilever benchmark with a 200 x 100 mesh. Clamp the left edge and apply downward pressure of 200 on a small patch at the lower-right corner. Use 30 percent material, run 100 optimization steps, and set the step size to 0.1.
\end{tcolorbox}

\subsection{3D cantilever beam}

\begin{tcolorbox}[breakable, colback=white, colframe=black, title=Clear Problem Specification: 3D Cantilever]
Create a 40 x 10 x 20 mesh. Fix the left side and apply downward pressure of 200 at the lower-right corner, where x = 36-40, z = 0, and y = 0-10. Set the volume fraction to 0.3, the iteration count to 100, and the step size to 0.1.
\end{tcolorbox}

\begin{tcolorbox}[breakable, colback=white, colframe=black, title=Ambiguous Problem Specification: 3D Cantilever]
Create a 3D cantilever benchmark with a 40 x 10 x 20 mesh. Fix the left face and apply downward pressure of 200 on a small lower-right end patch. Use 30 percent material, run 100 optimization steps, and set the step size to 0.1.
\end{tcolorbox}

\subsection{2D bridge structure}

\begin{tcolorbox}[breakable, colback=white, colframe=black, title=Clear Problem Specification: 2D Bridge]
Create a 200 x 100 mesh. Fix the lower-right corner (x = 190-200, y = 0) and the lower-left corner (x = 0-10, y = 0), and apply downward pressure of 1000 on the top (x = 0-200, y = 99.5-100). Set the elements on the top as a passive domain (x = 0-200, y = 98.5-100). Set the volume fraction to 0.2, the iteration count to 100, and the step size to 0.1. Set the filter factor to 2.
\end{tcolorbox}

\begin{tcolorbox}[breakable, colback=white, colframe=black, title=Ambiguous Problem Specification: 2D Bridge]
Create a 2D bridge benchmark with a 200 x 100 mesh. Support both lower corners and apply downward pressure of 1000 across the top edge. Keep the top strip as a passive domain. Use 20 percent material, run 100 optimization steps, set the step size to 0.1, and use a filter factor of 2.
\end{tcolorbox}

\subsection{3D bridge structure}

\begin{tcolorbox}[breakable, colback=white, colframe=black, title=Clear Problem Specification: 3D Bridge]
Create a 40 x 5 x 20 mesh. Fix the lower-right corner (x = 38-40, z = 0, and y = 0-5) and the lower-left corner (x = 0-2, z = 0, and y = 0-5), and apply downward pressure of 1000 on the top (x = -1-41, y = -1-6, and z = 19.5-20). Set the elements on the top as a passive domain (x = -1-41, y = -1-6, and z = 18.5-20). Set the volume fraction to 0.2, the iteration count to 100, and the step size to 0.1. Set the filter factor to 2.
\end{tcolorbox}

\begin{tcolorbox}[breakable, colback=white, colframe=black, title=Ambiguous Problem Specification: 3D Bridge]
Create a 3D bridge benchmark with a 40 x 5 x 20 mesh. Support both lower end corners and apply downward pressure of 1000 across the top surface. Keep the top layer as a passive domain. Use 20 percent material, run 100 optimization steps, set the step size to 0.1, and use a filter factor of 2.
\end{tcolorbox}

\subsection{2D thermal conduction}

\begin{tcolorbox}[breakable, colback=white, colframe=black, title=Clear Problem Specification: 2D Thermal Conduction]
Please run a 2D thermal conduction topology optimization. Use a square 2D domain from x = 0 to 5 and y = 0 to 5 with an element size of 0.1. Apply a uniform internal heat source over the whole domain. Use a fixed-temperature cold sink on the middle part of the left boundary: x = 0, y from 2.25 to 2.75, with temperature T = 0. The other three boundaries, x = 5, y = 0, and y = 5, should be insulated, so do not add heat flux boundary conditions. Set the volume fraction to 0.35, kmax to 1.0, kmin to 1e-3, and the heat source to 1.0. Run 80 steps.
\end{tcolorbox}

\begin{tcolorbox}[breakable, colback=white, colframe=black, title=Ambiguous Problem Specification: 2D Thermal Conduction]
Please run a 2D thermal conduction topology optimization. Use a square 2D domain from x = 0 to 5 and y = 0 to 5 with an element size of 0.1. Apply a uniform internal heat source over the whole domain. Use a fixed-temperature cold sink on the middle part of the left boundary with temperature T = 0. Leave the other three boundaries insulated. Set the volume fraction to 0.35, kmax to 1.0, kmin to 1e-3, and the heat source to 1.0. Run 80 steps.
\end{tcolorbox}

\subsection{3D multi-load}

\begin{tcolorbox}[breakable, colback=white, colframe=black, title=Clear Problem Specification: 3D Multi-load]
Create a 10 x 10 x 40 mesh. Fix the bottom (z = 0). Use two load cases: first, apply downward pressure of 1000 on the top (x = 0-10, y = 0-10, and z = 40); second, apply pressure of 100 to the right on the left surface (x = 0, y = 0-10, and z = 0-40). Set the elements on the top (x = 0-10, y = 0-10, and z = 38-40) and the left surface (x = 0-1, y = 0-10, and z = 0-40) as passive domains. Set the volume fraction to 0.2, the iteration count to 100, and the step size to 0.1. Set the filter factor to 2.
\end{tcolorbox}

\begin{tcolorbox}[breakable, colback=white, colframe=black, title=Ambiguous Problem Specification: 3D Multi-load]
Create a 3D multi-load benchmark with a 10 x 10 x 40 mesh. Fix the bottom face. Use two load cases: one with downward pressure of 1000 on the top surface and one with horizontal pressure of 100 pushing to the right on the left surface. Keep the top region and the left surface region as passive domains. Use 20 percent material, run 100 optimization steps, set the step size to 0.1, and use a filter factor of 2.
\end{tcolorbox}




\end{appendices}


\input{sn-article.bbl}


\end{document}